\documentclass[prd,aps,floats,preprintnumbers,preprint]{revtex4}
\usepackage{graphicx}
\usepackage[english]{babel} % grammar package
\usepackage{braket}%Dirac notation
\newcommand{\be}{\begin{equation}}
\newcommand{\ee}{\end{equation}}
\newcommand{\bea}{\begin{eqnarray}}
\newcommand{\eea}{\end{eqnarray}}

\newcommand{\eqn}[1]{(\ref{#1})}

\usepackage{morefloats} %When there are many figs.~
\usepackage[titletoc,title]{appendix}
\usepackage{color}
\begin{document}

\title{Effects of  discontinuities of the derivatives of the inflaton potential}

\author{Alexander Gallego Cadavid$^{3,2}$, Antonio Enea Romano$^{3,1,2}$}
\affiliation
{
${}^{1}$Department of Physics, University of Crete, 71003 Heraklion,Greece \\
${}^{2}$Yukawa Institute for Theoretical Physics, Kyoto University, Japan\\
${}^{3}$Instituto de Fisica, Universidad de Antioquia, A.A.1226, Medellin, Colombia
}

\begin{abstract}
We study  the effects of a class of features of the inflaton potential, corresponding to discontinuties in its derivatives. We perform fully numerical calculations and derive analytical approximations for the curvature pertubations spectrum and the bispectrum which are in good agreement with the numerical results. 

The spectrum of primordial perturbations has  oscillations around the scale $k_0$ which leaves the horizon at the time $\tau_0$ when the feature occurs, with the amplitude and phase of the oscillations determined by the size and the order of the discontinuity. The large scale bispectrum in the squeezed and equilateral limits have a very similar form and are linearly suppressed. Both in the squeezed and equilateral small scale limit the bispectrum has an oscillatory behavior whose phase depends on the parameters determining the discontinuity, and whose amplitude is inversely proportional to the scale. Given the generality of this class of features they could be used to model or classify phenomenologically different types of non Gaussian features encountered in observational data such as the cosmic microwave background radiation or large scale structure. 
\end{abstract}

\maketitle

\section{Introduction}
In the last few decades the outstanding advances in observational cosmology have allowed for the first time to test theoretical cosmological  models  \cite{et, wmapcpr, pxvi,inflation2014}. Among the most important sources of cosmological observational data we can mention the Sloan Digital Sky Survey (SDSS), the Wilkinson Microwave Anisotropy Probe (WMAP), and the Planck mission, and other ground-based and sub-orbital experiments \cite{gbe1,gbe2}. According to the standard cosmological model the cosmic microwave background (CMB) radiation consists of photons which decoupled from the primordial plasma at the time when protons and electrons combined to form neutral light atoms. Although this radiation is extremely isotropic there are small fluctuations in the temperature  of the order of $\Delta T/T \sim 10^{-5}$. And since the CMB radiation was emitted at a redshift of about $1100$ it provides a unique window on the early universe \cite{wmapfmr, pxxii, xc}. 

Inflation theory \cite{anewtype} explains the anisotropies of the CMB temperature as the consequence of  primordial curvature perturbations whose statistical properties can be described by the n-points correlation functions. If the perturbations followed a perfectly Gaussian distribution the two points correlation function would be enough, but even the most recent observations are  compatible with some non Gaussianity corresponding to $f_{NL}^{local}=2.5\pm5.7$ and $f_{NL}^{equil}=-16\pm70$~\cite{pxxii, pxxiv}, motivating the theoretical study of the conditions which could have generated it.
Some recent developments in the study of models which could generate non Gaussianity and on their detection can be found for example in \cite{Hazra:2012yn,Dorn:2014kga,bingo,numerical2013}.

The theoretical study of the effects of features of the inflaton potential was started in the seminal works of Starobinsky \cite{starobinsky}, and once CMB observational data became available it was shown that features can  be used to model the glitches of the power spectrum \cite{constraints1,constraints2}.
%  see also the review in astro-ph/9811360 where physical mechanisms leading to such a feature in the inflaton potential are discussed, too. 
% The model (5) is a particular case of the model considered in arXiv:1404.0360 for p=2, see also arXiv:1405.2012.
Some other interesting studies and reviews in this area can be found for example in \cite{Starobinsky:1998mj,Joy:2007na,Joy:2008qd,Mortonson:2009qv}.
In this paper we focus on the effects of features of the inflaton potential on the primordial curvature perturbations, considering a class corresponding to a discontinuity in the derivatives of the potential. Our model is a generalization of other features which have been studied earlier such as the Starobinsky model or the mass step \cite{aer}. These kinds of features could have arisen through different mechanisms such as for example particle production \cite{pp}, or phase transitions \cite{Adams2}, but in this paper we study their effects from a purely phenomenological point of view, without investigating their  fundamental origin.

There is also an important observational motivation for studying this kind of potentials: recent analyses of CMB observations based on cubic Hermite interpolating polynomials (PCHIP) for the primordial curvature perturbations spectrum \cite{Gariazzo:2014dla} have in fact shown some evidence for a  feature around the wave number  $k=0.002$ Mpc${}^{-1}$, which is in good qualitative agreement with  the results of our calculations for some of the potentials we consider.

The paper is organized as following: first we define the features, then we give both a numerical and analytical solution for the background, and finally provide both numerical and analytical calculations of the spectrum and the bispectrum, giving details of the squeeze and equilateral limit and show the effects of varying the different parameters defining the feature, i.e., its amplitude and the order  $n$ of the discontinuous derivatives. 
%*********************************************************************************************************************************************
\section{Inflation}\label{inflation}

We consider inflationary models with a single scalar field and a standard kinetic term according to the action \cite{linf, inf}
\begin{equation}\label{action1}
  S = \int d^4x \sqrt{-g} \left[ \frac{1}{2} M^2_{Pl} R  - \frac{1}{2}  g^{\mu \nu} \partial_\mu \phi \partial_\nu \phi -V(\phi)
\right],
\end{equation}
where $ M_{Pl} = (8\pi G)^{-1/2}$ is the reduced Planck mass. Varying the action with respect to the metric tensor 
and the scalar field we get Friedmann equation and the equation of motion of the inflaton
\begin{equation}\label{ema}
  H^2 \equiv \left(\frac{\dot a}{a}\right)^2= \frac{1}{3 M^2_{Pl}}\left( \frac{1}{2} \dot \phi^2 + V(\phi) \right),
\end{equation}
\begin{equation}\label{emphi}
  \ddot \phi + 3H\dot \phi + \partial_{\phi}V = 0,
\end{equation}
where dots and $\partial_{\phi}$ indicate derivatives with respect to time and scalar field respectively and $H$ is the Hubble parameter. We adopt the following definitions of the slow roll parameters
\bea \label{slowroll}
  \epsilon \equiv -\frac{\dot H}{H^2} \,\,\,\, , \,\,\,\, \eta \equiv \frac{\dot \epsilon}{\epsilon H}.
\eea
%*******************************************************************************************************************************************
\section{The model}\label{m}
We consider a single scalar field $\phi$ with potential
\begin{equation}\label{pot}
V(\phi)= \left\{
\begin{array}{lr}
V_{b} + \frac{1}{2}m^2 \phi^2, & \phi > \phi_0 \\
V_{a} + \frac{1}{2}m^2 \phi^2 + \lambda \Delta \phi, \, & \phi < \phi_0
\end{array}
\right.
\end{equation}
where
\bea
\Delta \phi \equiv \phi^n.
\eea
where $V_a$ and $V_b$ are different in order to ensure the continuity of the potential at $\phi_0$.

The value of $\phi_0$ determines the scale at which the effects of the feature appear in the power spectrum and the bispectrum of curvature perturbations, and as such it is a free parameter which can be fixed phenomenologically based on experimental data.
It is in fact determining the value of conformal time when $\phi(\tau_0)=\phi_0$,
and as it will be shown in the following sections the features in the spectrum and bispectrum appear around the scale $k_0=-1/\tau_0$ which is leaving the horizon at that time.

The potential has a discontinuity in the derivatives at $\phi_0$, which is the cause of the temporary slow roll regime violation which produce the non Gaussian features.
The continuity condition for the potential at $\tau_0$ gives 
\bea
V_{a}= V_{b}-\lambda \phi_0^n \,. 
\eea
In this paper we study potentials dominated by the vacuum energies $V_{b}$ and $V_{a}$ before and after the feature.
The potential in eq.~(\ref{pot}) is similar to the one studied in \cite{whipped,Bousso:2013uia}, but it only coincide with it in the special case $(\phi_0=0,p=2)$. Another important difference is that inflation in our model is driven by the dominating vacuum energy term $V_a$.

\section{Analytic solution of the background equations}\label{asem}
The Friedmann equation and the equation of motion for the inflaton in terms of conformal time $\tau$ take the form
\begin{equation}\label{sfe}
  H^2  \equiv \left(\frac{a'}{a^2}\right)^2 = \frac{1}{3 M^2_{Pl}} \left( \frac{1}{2} \frac{{\phi'}^{2}}{a^2} + V(\phi)\right),
\end{equation}
\begin{equation}\label{be}
  \phi'' + 2\frac{a'}{a} \phi' + a^2 \partial_{\phi}V = 0,
\end{equation}
where primes indicate derivatives with respect to conformal time. 

Since $V(\phi)$ is dominated by the vacuum energy we can use the De Sitter approximation in which $H$ is a constant and the scale factor is given by
\begin{equation}
 a(\tau)= \frac{-1}{H \tau}\,.
\end{equation}
Before the feature the equation of motion of the inflaton is
\begin{equation}\label{beb}
  \phi'' + 2\frac{a'}{a} \phi' + a^2 m^2 \phi = 0,
\end{equation}
which has the solution
\begin{equation}\label{bsb}
  \phi_b(\tau)= \phi_b^+ a(\tau)^{\lambda^+} + \phi_b^- a(\tau)^{\lambda^-}\, ,
\end{equation}
where
\begin{equation}\label{lambdaplusminus}
  \lambda^{\pm}=\frac{3}{2}\left( -1\pm \sqrt{1- \left(\frac{2 m}{3 H}\right)^2}\right) \,.
\end{equation}
The slow roll regime corresponds to $\phi_b^-=0$.
After the feature the equation of motion of the inflaton becomes 
\begin{equation}\label{bea}
\phi'' + 2\frac{a'}{a} \phi' + a^2 \left( m^2 \phi + \lambda n \phi ^{n-1} \right) = 0,
\end{equation}
In order to find an analytical solution we can expand the last term in  eq.~\eqn{bea} to second order in conformal time around $\tau_0$
\bea \label{bea2}
  \phi'' + 2\frac{a'}{a} \phi' + a^2 \left\{ m^2 \phi + n \lambda  \phi _0{}^{n-1} \left[ 1 + (n-1) \frac{\phi'(\tau_0)}{\phi_0} (\tau -\tau_0) + \right. \right. \\ \nonumber
  \left. \left. \frac{(n-1)}{2} \left( (n-2) \frac{\phi'(\tau_0)^2}{\phi_0^2} + \frac{\phi''(\tau_0)}{\phi_0} \right) (\tau -\tau_0)^2 \right] \right\}= 0.
\eea
From now on quantities evaluated at $\tau_0$ are denoted by the subscript $0$. From eq.~\eqn{bsb} we have an analytical expression for the first and second derivative of the field at $\tau_0$ 
\bea \label{phiderivatives}
\phi'_0 =\lambda^+ \phi_b^+ a_0^{\lambda^+-1} a'_0 =\lambda^+ \frac{a'_0}{a_0} \phi_0 =\lambda^+ a_0 H \phi_0, \\
\phi''_0 =\lambda^+ H (a_0 \phi'_0 + a'_0 \phi_0 ) = \lambda^+ H(\lambda^+ a_0^2 H \phi_0 + a_0^2 H \phi_0)\approx \lambda^+ a_0^2 H^2 \phi_0 \, ,
\eea
where we have only kept terms linear in $\lambda^+$ because during slow regime higher order terms can be safely neglected according to
\be
|\lambda^+| \approx \frac{1}{3}\frac{m^2}{H^2} \ll 1.
\ee
Assuming the same slow roll regime condition we can expand eq.~(\ref{bea2}) to linear order in $\lambda^+$ and get
\bea
\phi'' + 2\frac{a'}{a} \phi' + a^2 \left\{ m^2 \phi + n \lambda  \phi _0{}^{n-2} \left[ \phi _0+ (n-1)\phi'_0 (\tau -\tau_0) + \frac{1}{2}(n-1) \phi''_0 (\tau -\tau_0)^2 \right] \right\}= 0,
\eea
which admits an analytical solution of the form
\begin{equation}\label{bsa}%\textsc{e}
  \phi_a(\tau)= \phi_{a}^{{}_{(0)}} +\phi_{a}^{{}_{(1)}} (\tau-\tau_0)+\phi_{a}^{{}_{(2)}} (\tau-\tau_0)^2 + 
  \phi_a^+ a(\tau)^{\lambda^+} + \phi_a^- a(\tau)^{\lambda^-},
\end{equation}
where
\bea
  \phi_{a}^{{}_{(0)}} = \frac{-n\lambda  \phi _{0}^{n-2}}{m^2 \left( m^2-2H^2 \right)} \left[ (m^2-2 H^2) \phi _{0} + 2(n-1) H^2\phi _{0}'\tau _0- (n-1)H^2 \phi _{0}''\tau _0{}^2 \right] \, , \\
  \phi_{a}^{{}_{(1)}} = \frac {-n(n-1)\lambda  \phi _{0} {}^{n - 2}} {\left (m^2-2 H^2 \right)} \phi _{0}'\, , \\ 
  \phi_{a}^{{}_{(2)}} = \frac{-n(n - 1)\lambda \phi_{0} {}^{n - 2}} {2\left (m^2 -2 H^2 \right)}  \phi _{0}'' \, .
\eea
The constants of integration $\phi_{a}^{\pm}$ are determined by imposing the continuity conditions for  $\phi$ and $\phi'$ at $\tau_0$ which give
\bea
  \phi_{a}^{\pm} = \frac{\pm1}{a(\tau_0){}^{\lambda^{\pm} } (\lambda^--\lambda^+) } \left\{ \lambda^\mp \phi _{0}+ \phi_{0}'\tau_0+\frac{n \lambda  \phi_{0}{}^{n-2}}{m^2}
	\right. \\ \nonumber
	\left. \times \left[ \lambda^\mp \phi_{0}+ \frac{(n-1)}{(m^2-2 H^2)} \left( (m^2+2 H^2 \lambda^\mp) \phi_{0}'\tau_0 -\lambda^\mp H^2 \phi_{0}''\tau_0^2 \right)\right] \right\}.
\eea
We can also find an analytical approximation for the slow roll parameters after the feature by substituting the eq.~(\ref{bsa}) in eq.~(\ref{slowroll})
%\begin{equation}\label{as}
%  \phi(\tau)= \phi_b(\tau) + \phi_a(\tau) \theta(\phi_0-\phi),
%\end{equation}
%where $\phi_{b}$ and  $\phi_{a}$ are defined in eq.~\eqn{bsb} and eq.~\eqn{bsa}, and $\theta$ %denotes the Heaviside step function. The analytical expressions for the slow-roll parameters%after the feature in terms of conformal time are
\bea\label{slowrollapprox}
  \epsilon_a(\tau)\approx \frac{1}{2}\left[\lambda^+ \phi_a^+ a(\tau)^{\lambda^+}+ \lambda^- \phi_a^- a(\tau)^{\lambda^-} \right]^2 \, , \\ \nonumber
  \eta_a(\tau)\approx 2 \frac{ (\lambda^+)^2 \phi_a^+ a(\tau)^{\lambda^+}+ (\lambda^-)^2\phi_a^- a(\tau)^{\lambda^-} }{
		\lambda^+ \phi_a^+ a(\tau)^{\lambda^+} + \lambda^- \phi_a^- a(\tau)^{\lambda^-} }.
\eea

\section{Numerical  solution of the background equations}
\begin{figure}[ht]
 \begin{minipage}{.45\textwidth}
  \includegraphics[scale=0.6]{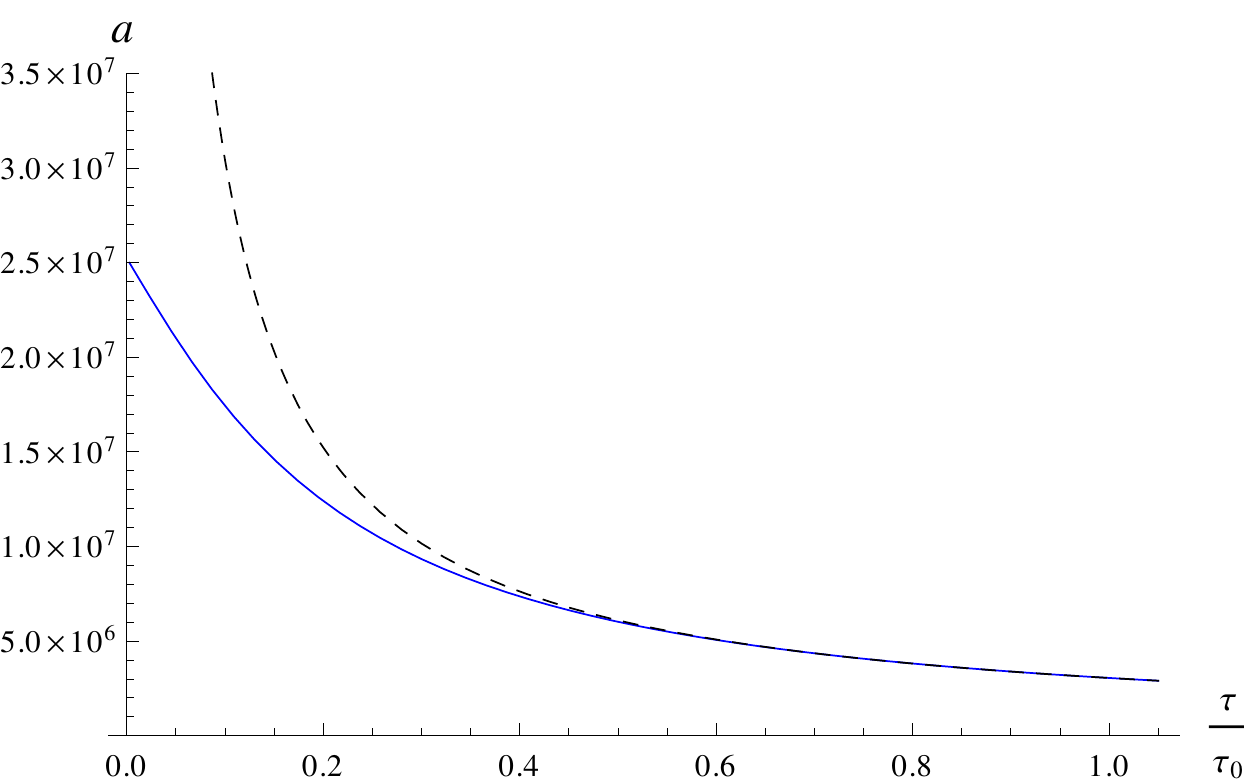}
  \end{minipage}
 \begin{minipage}{.45\textwidth}
  \includegraphics[scale=0.6]{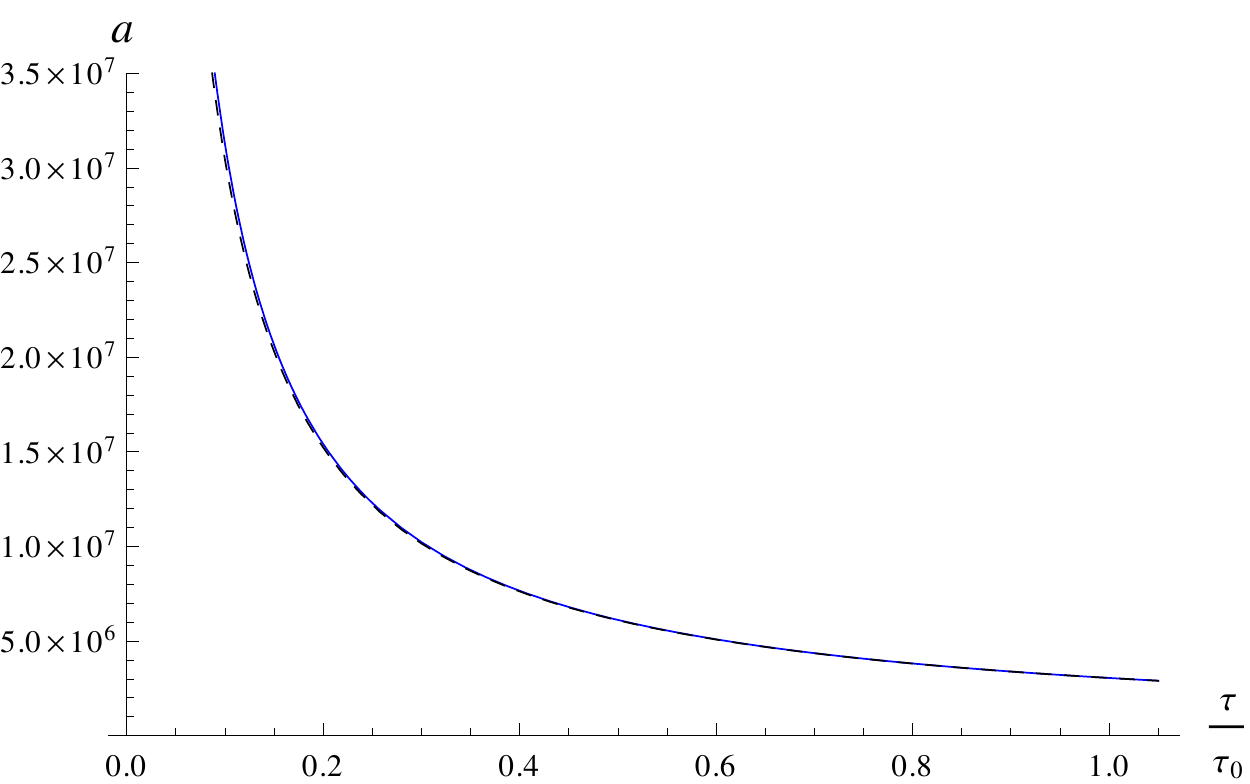}
 \end{minipage}
 % adesitterviolation.pdf: 0x0 pixel, 300dpi, 0.00x0.00 cm, bb=0 0 360 213
 \caption{Numerical (blue) and analytical (dashed black) evolution of the scale factor as a function of conformal time is plotted for $n=3$. On the left we choose $\lambda =7.2\times 10^{-16}$ and on the right $\lambda =2.4\times 10^{-18}$. As it can be seen in the left plot the De Sitter approximation is not valid at late times when $\lambda \phi^n > V_0$, so that a full numerical integration of the background equations is required, while from the right plot we can see that when $\lambda \phi^n < V_0$ the De Sitter approximation is quite accurate.}
 \label{adesitterviolation}
\end{figure}

The background evolution can be obtaining by solving the system of coupled differential equations for $a(\tau)$ and $\phi(\tau)$, or alternatively $H(\tau)$ and $\phi(\tau)$.
In the numerical integration we chose the following value for the different parameters defining the model
\begin{equation}\label{parameters}
  m\approx 6\times10^{-9}M_{Pl}, \,\,\,\, H=3.3\times 10^{-7}M_{Pl},\,\,\,\, \phi_b^+=10M_{Pl}.
\end{equation}
This choice of the parameters is made in order to satisfy the Planck normalization on small scales.
If the term $\lambda \Delta\phi$ is of the order of the vacuum energy $V_{b}$ then from eq.~\eqn{ema} we have
\begin{equation}\label{desitterlimit}
 H^2  \approx \frac{1}{3 M^2_{Pl}} V(\phi) \approx\frac{1}{3 M^2_{Pl}} \left( V_{b} + \lambda \phi^n\right) \,,
\end{equation}
implying that in this case the De Sitter approximation used to obtain an analytical solution in the previous section is not valid as shown in fig.~(\ref{adesitterviolation}) and the numerical integration is necessary to obtain reliable results. 
As stated previously, we will focus on vacuum energy dominated models for which the De Sitter approximation is valid, so fig.~(\ref{adesitterviolation}) is given only to show the limits of its validity, but in all the cases we consider it turns out to be quite accurate as shown in the figures comparing analytical results, based on the De sitter approximation, and  to numerical results, which take into account the small variation of the Hubble parameter.
We adopt a system of units in which $M_{Pl}=1$. 

\begin{figure}
 \begin{minipage}{.45\textwidth}
  \includegraphics[scale=0.6]{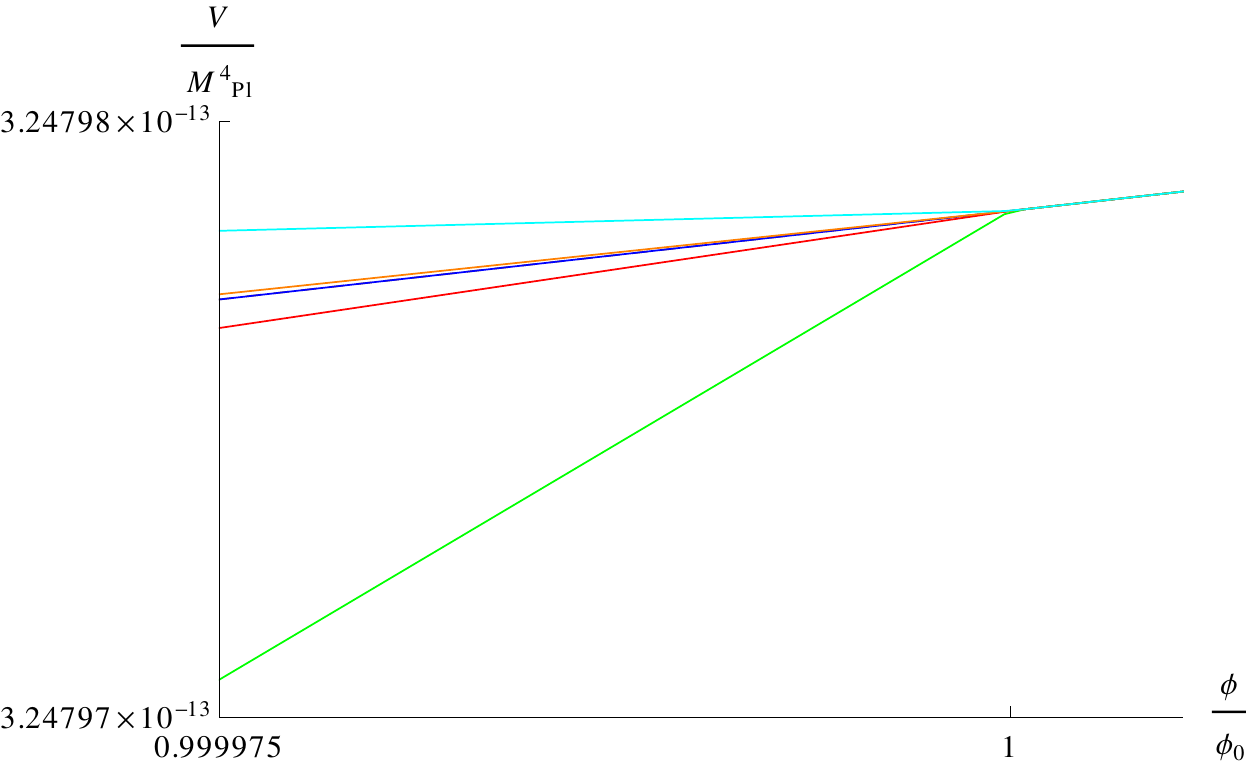}
  \end{minipage}
 \begin{minipage}{.45\textwidth}
  \includegraphics[scale=0.6]{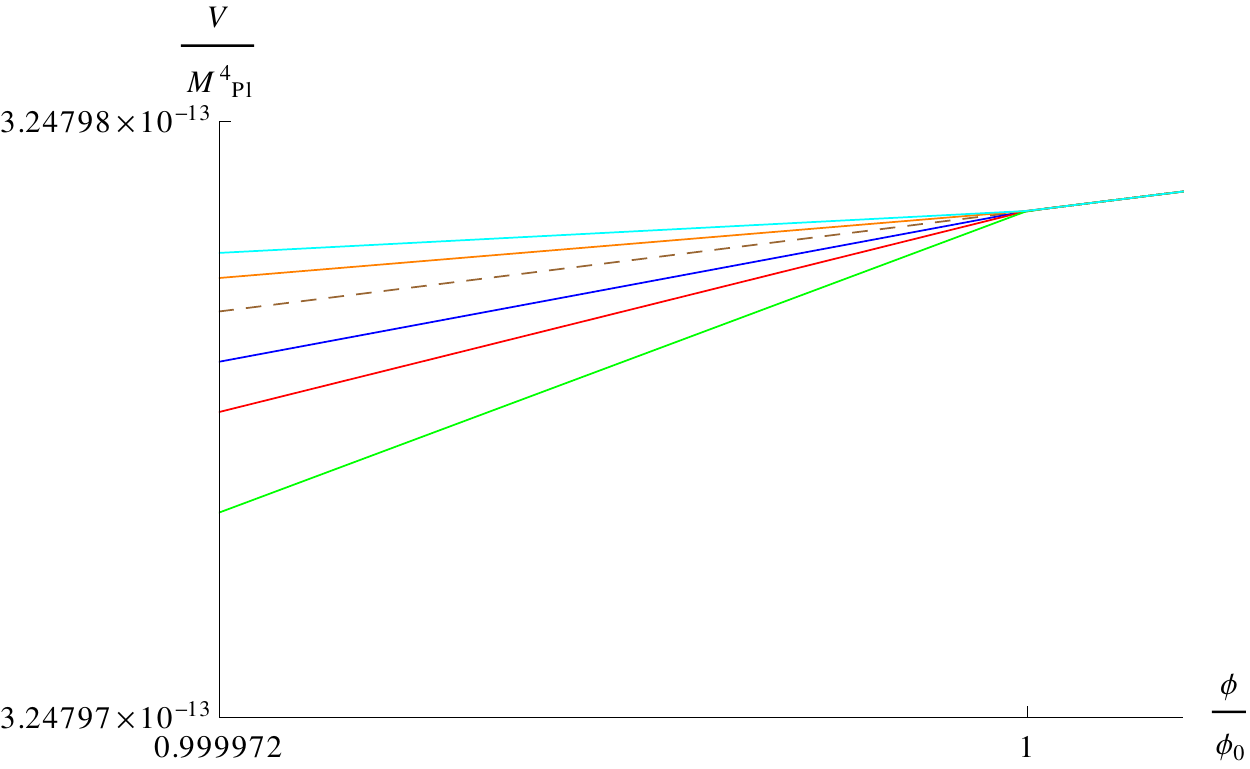}
 \end{minipage}
 \caption{On the left the potential $V$ is plotted  as a function of the field $\phi$ for $\lambda=3.9\times10^{-19}$ and $n=2/3$ (blue), $n=3$ (red), and $n=4$ (green) and for $\lambda=-7\times10^{-20}$ and $n=3$ (orange) and $n=4$ (cyan). On the right the potential $V$ is plotted  as a function of the field $\phi$ for $n=3$ and $\lambda=6.0\times10^{-19}$ (blue), $\lambda=1.2\times10^{-18}$ (red), $\lambda=2.4\times10^{-18}$ (green), $\lambda=-4\times10^{-19}$ (orange), and $\lambda=-7\times10^{-19}$ (cyan). The dashed brown lines correspond to the potential with no feature.}
\label{Vn123}
\end{figure}

\begin{figure}
 \begin{minipage}{.45\textwidth}
  \includegraphics[scale=0.6]{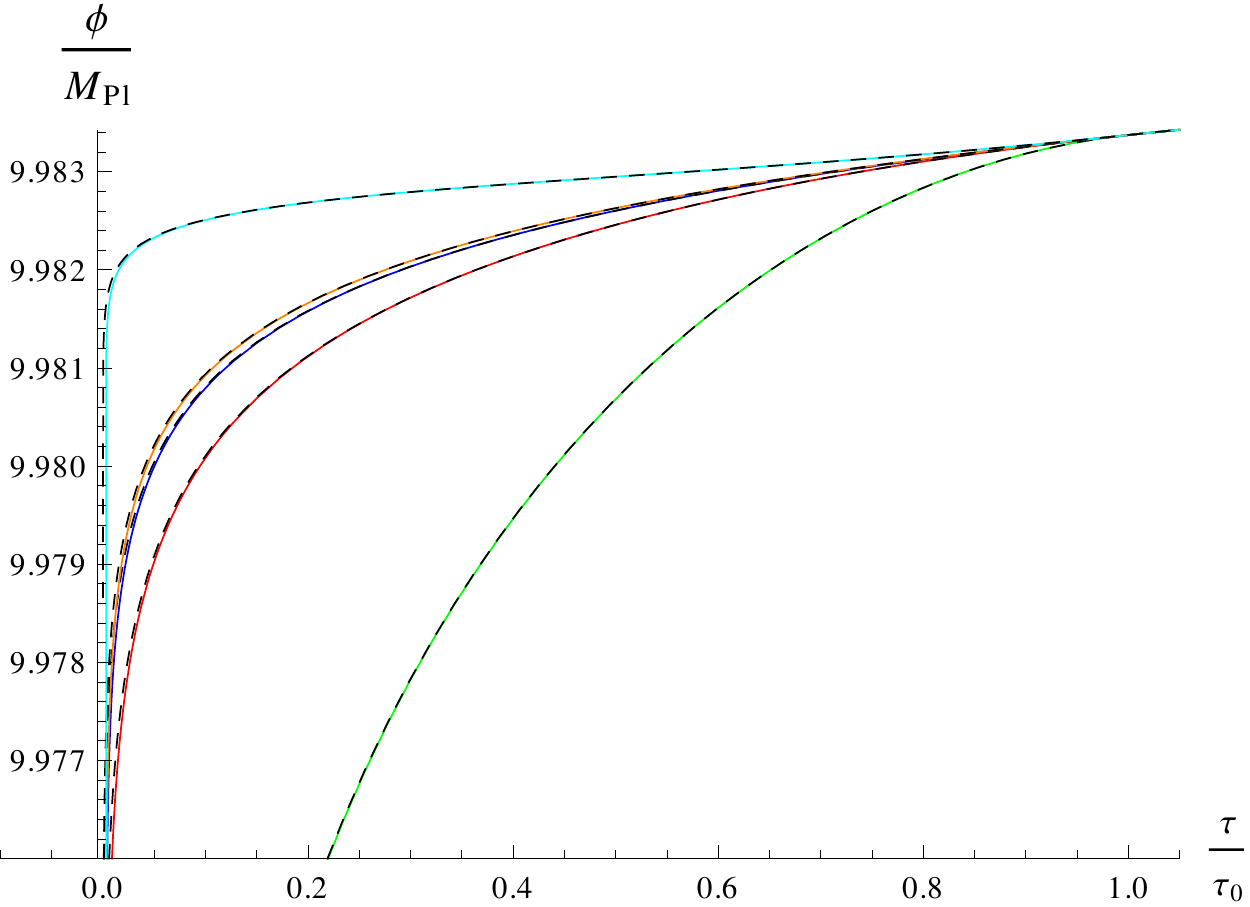}
  \end{minipage}
 \begin{minipage}{.45\textwidth}
  \includegraphics[scale=0.6]{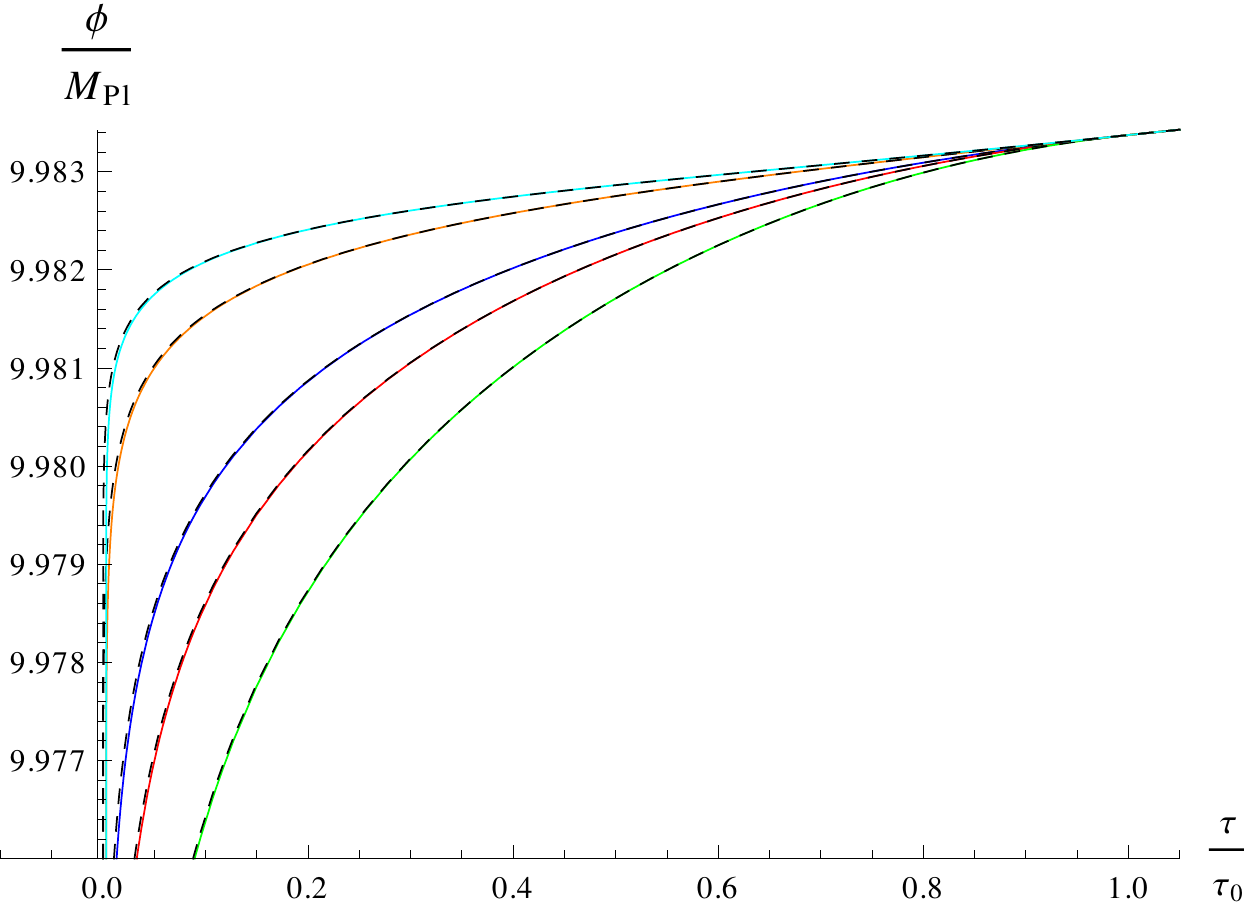}
 \end{minipage}
 \caption{On the left the numerically computed  $\phi$ is plotted as a function of conformal time for  $\lambda=3.9\times10^{-19}$ and $n=2/3$ (blue), $n=3$ (red), and $n=4$ (green) and for $\lambda=-7\times10^{-20}$ and $n=3$ (orange) and $n=4$ (cyan). On the right the numerically computed $\phi$ is plotted as a function of conformal time for $n=3$ and $\lambda=6.0\times10^{-19}$ (blue), $\lambda=1.2\times10^{-18}$ (red), $\lambda=2.4\times10^{-18}$ (green), $\lambda=-4\times10^{-19}$ (orange), and $\lambda=-7\times10^{-19}$ (cyan). The dashed black lines correspond to the analytical approximation.}
\label{Phin123}
\end{figure}

\begin{figure}
 \begin{minipage}{.45\textwidth}
  \includegraphics[scale=0.6]{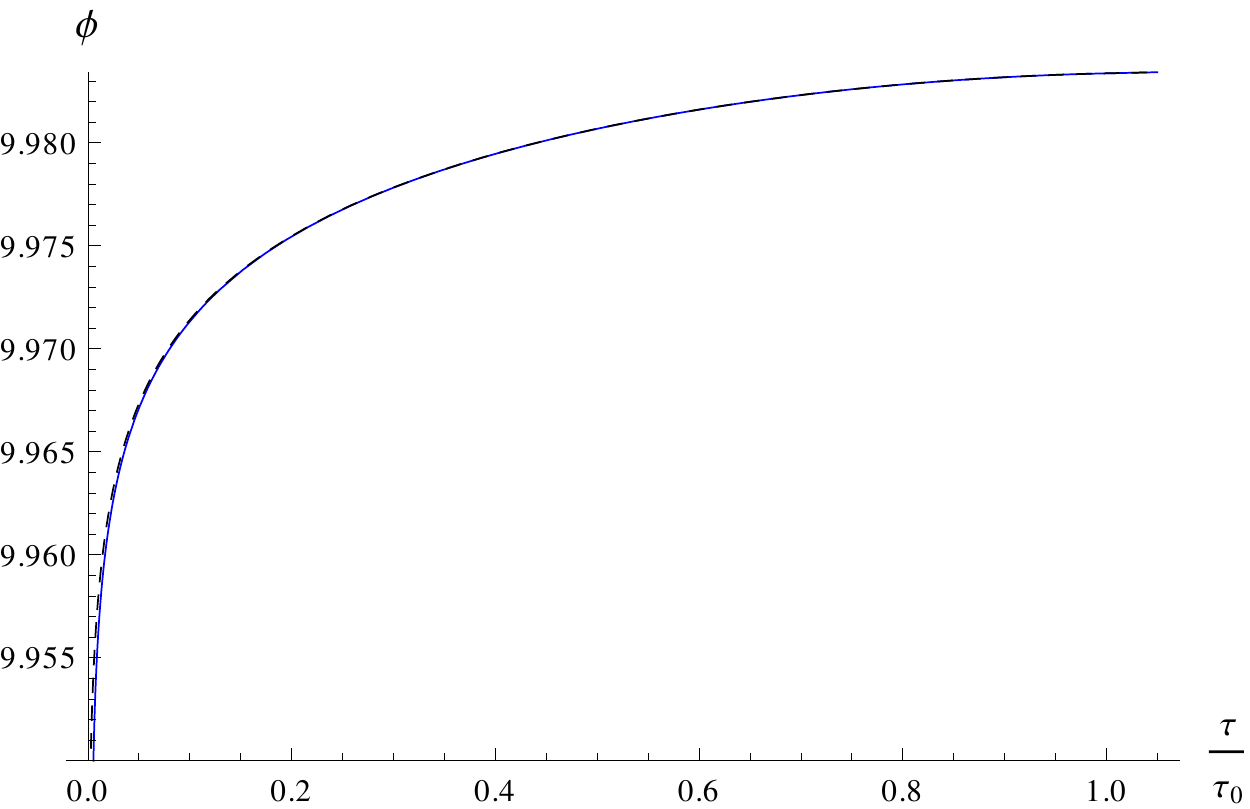}
  \end{minipage}
 \begin{minipage}{.45\textwidth}
  \includegraphics[scale=0.6]{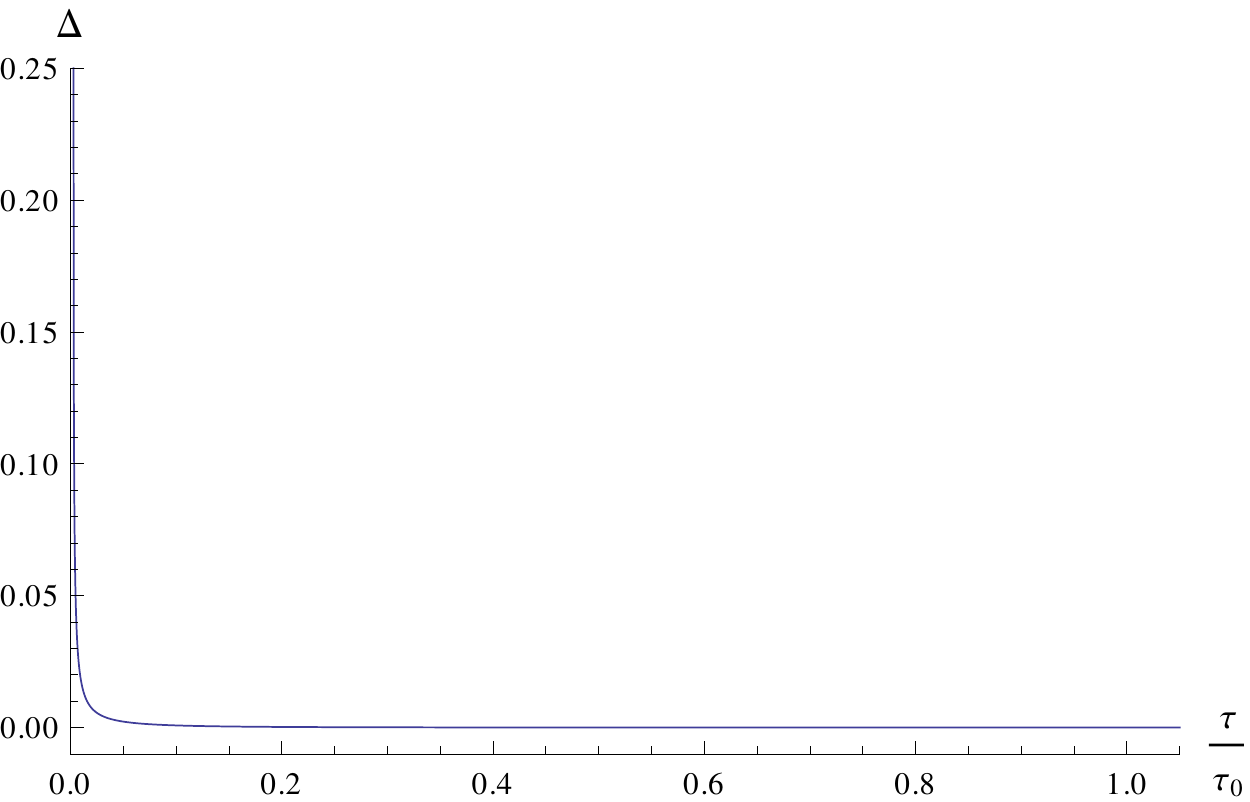}
 \end{minipage}
 \caption{On the left  $\phi$ is plotted in terms of conformal time for $n=4$ and $\lambda=3.9\times10^{-19}$. The blue and dashed-black lines are the numerical and analytical results, respectively. On the right the relative percentage error $\Delta=100\frac{(\phi^{num}-\phi^{an})}{\phi^{num}}$ between the numerical and analytical solutions for $\phi$ is plotted for the same values of $n$ and $\lambda$.}
\label{Pphiplot}
\end{figure}

\begin{figure}
 \begin{minipage}{.45\textwidth}
  \includegraphics[scale=0.6]{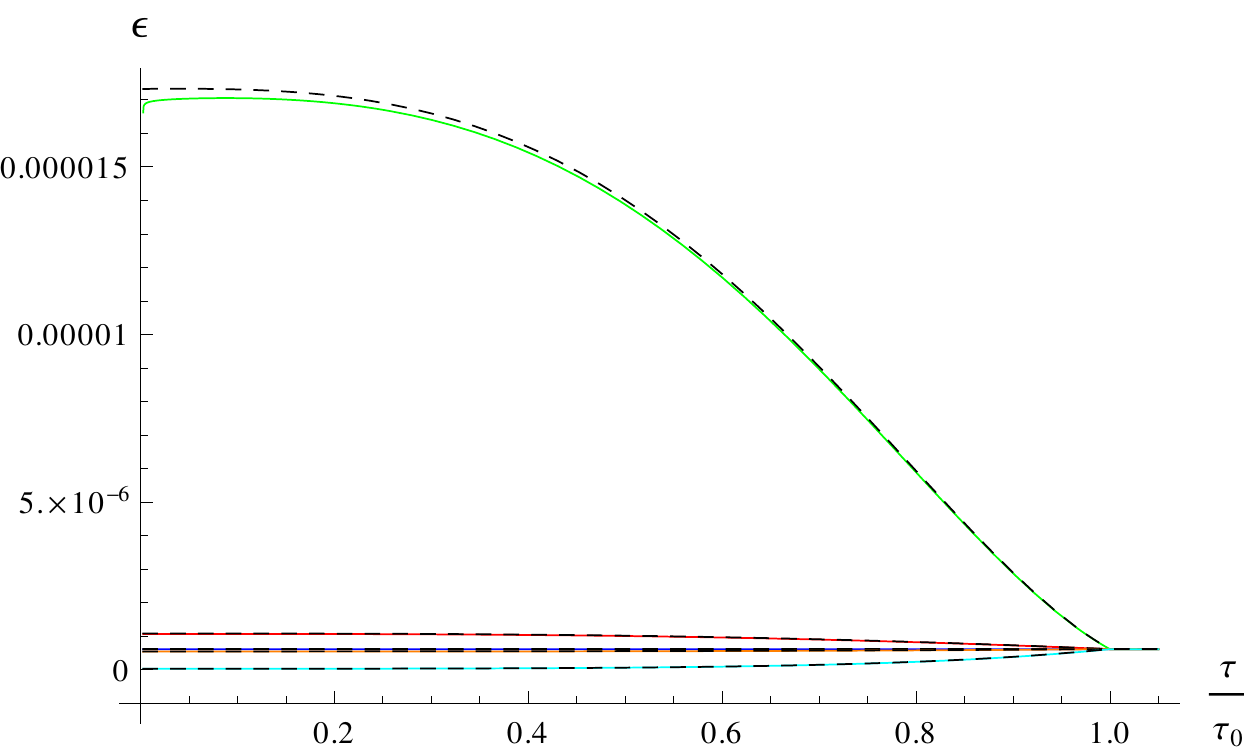}
  \end{minipage}
 \begin{minipage}{.45\textwidth}
  \includegraphics[scale=0.6]{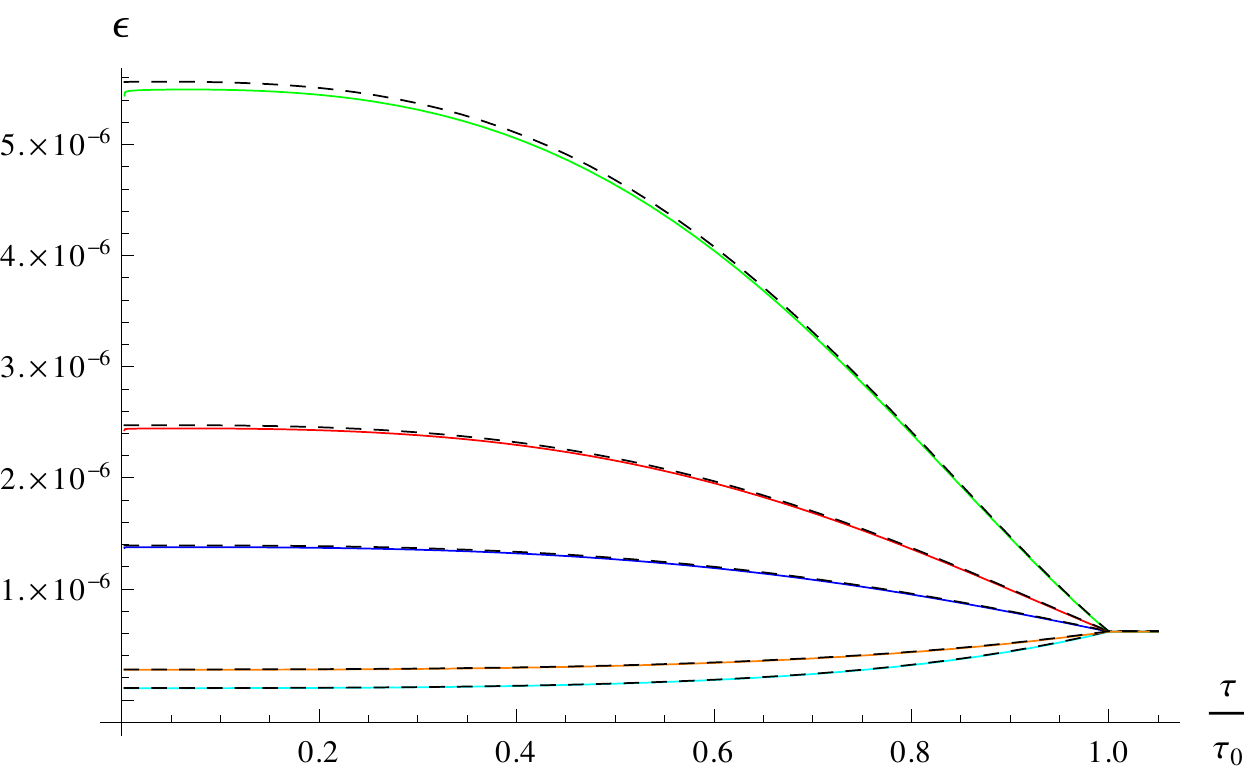}
 \end{minipage}
 \caption{On the left the numerically computed $\epsilon$ is plotted for $\lambda=3.9\times10^{-19}$ and $n=2/3$ (blue), $n=3$ (red), and $n=4$ (green) and for $\lambda=-7\times10^{-20}$ and $n=3$ (orange) and $n=4$ (cyan). On the right the numerically computed $\epsilon$ is plotted for $n=3$ and $\lambda=6.0\times10^{-19}$ (blue), $\lambda=1.2\times10^{-18}$ (red), $\lambda=2.4\times10^{-18}$ (green), $\lambda=-4\times10^{-19}$ (orange), and $\lambda=-7\times10^{-19}$ (cyan). The dashed black lines correspond to the analytical approximation.}
\label{Epsn123}
\end{figure}

\begin{figure}
 \begin{minipage}{.45\textwidth}
  \includegraphics[scale=0.6]{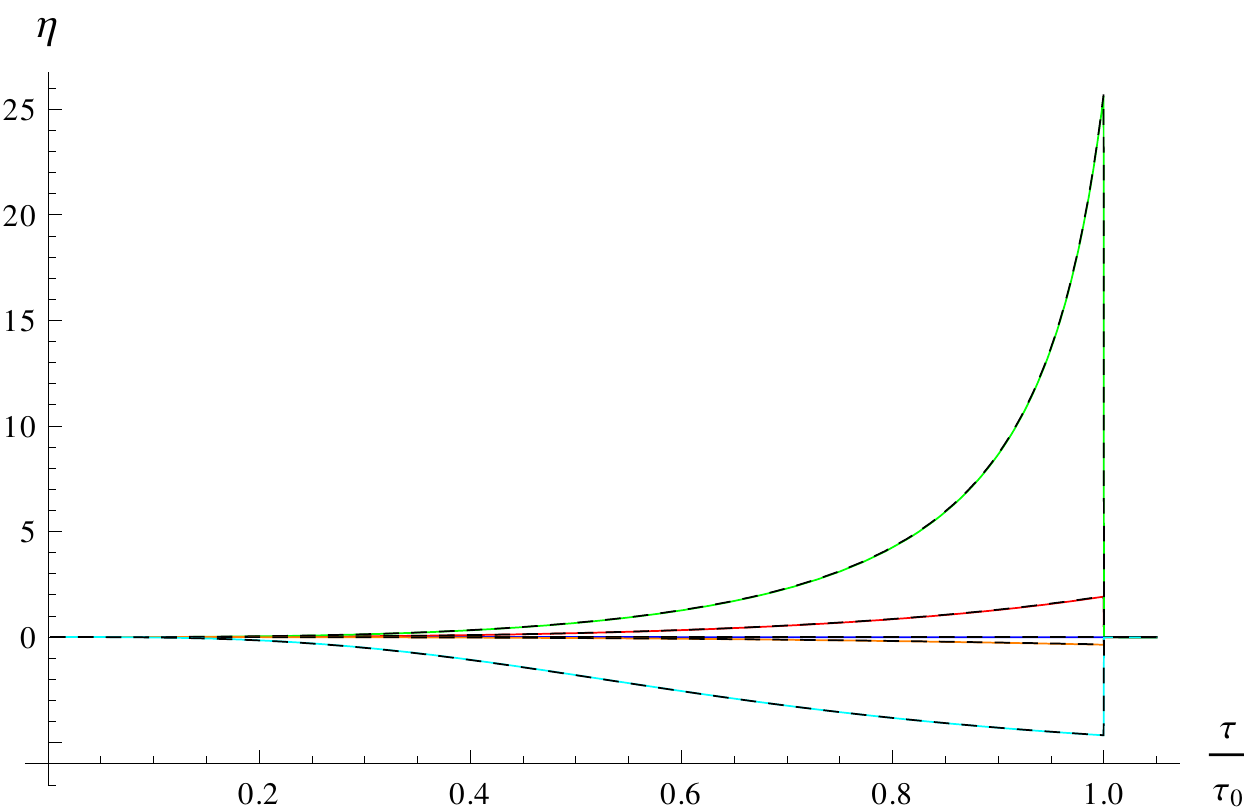}
  \end{minipage}
 \begin{minipage}{.45\textwidth}
  \includegraphics[scale=0.6]{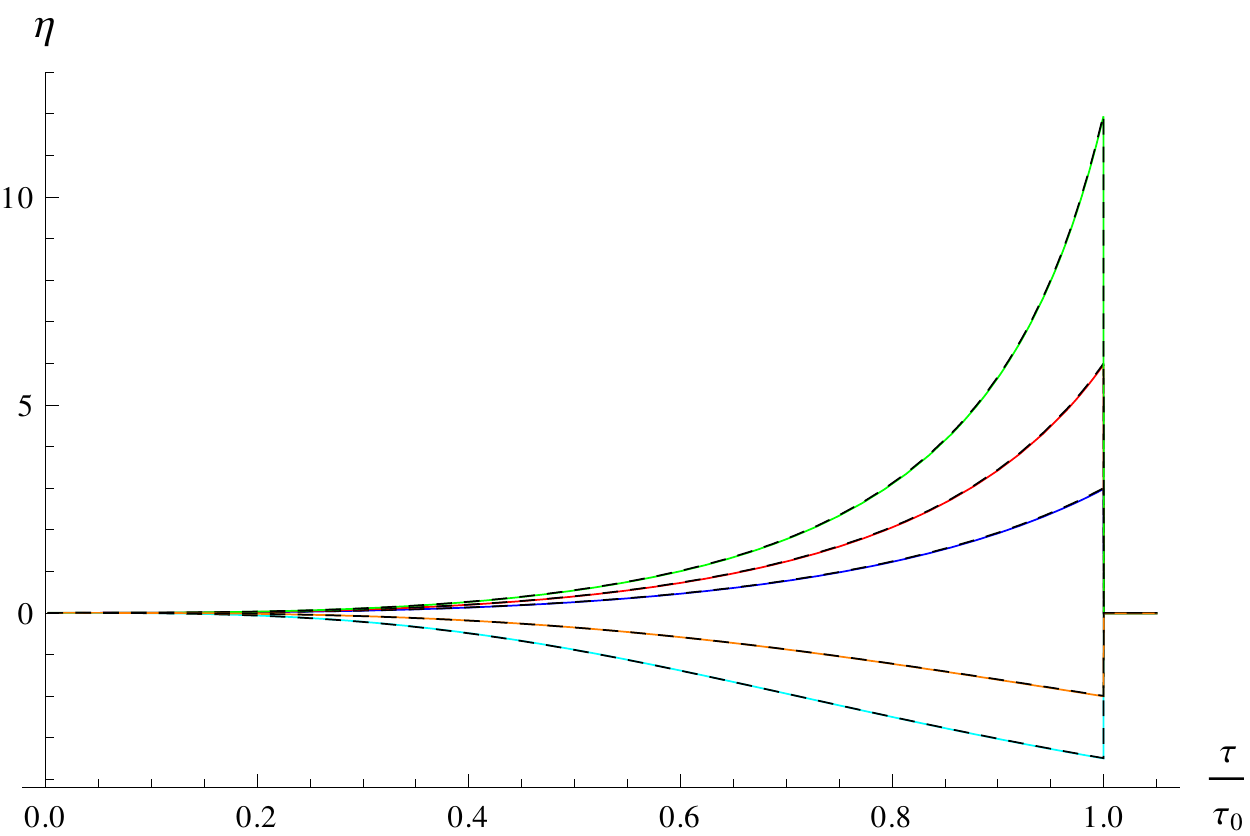}
 \end{minipage}
 \caption{On the left the numerically computed $\eta$ is plotted for $\lambda=3.9\times10^{-19}$ and $n=2/3$ (blue), $n=3$ (red), and $n=4$ (green) and for $\lambda=-7\times10^{-20}$ and $n=3$ (orange) and $n=4$ (cyan). On the right the numerically computed $\eta$ is plotted for $n=3$ and $\lambda=6.0\times10^{-19}$ (blue), $\lambda=1.2\times10^{-18}$ (red), $\lambda=2.4\times10^{-18}$ (green), $\lambda=-4\times10^{-19}$ (orange), and $\lambda=-7\times10^{-19}$ (cyan). The dashed black lines correspond to the analytical approximation.}
\label{Etan123}
\end{figure}

The potential as a function of the field is shown in fig.~(\ref{Vn123}) for different types of features. The effects of the features on the scalar field $\phi$  and on the slow roll parameters are shown in figs.~(\ref{Phin123}-\ref{Etan123}). When $|\lambda|$ is kept constant, larger values of $n$  tend to produce larger variations of both $\epsilon$ and $\eta$, while when  $n$ is kept constant, larger values of $|\lambda|$ tend to produce larger variations of both $\epsilon$ and $\eta$. As it can be seen in figs.~(\ref{Phin123}-\ref{Pphiplot}) the numerical and analytical solution for the scalar field are in  a good agreement. The analytical approximation is also good for the slow roll parameters as shown in figs.~(\ref{Epsn123}-\ref{Etan123}). We can conclude  that eq.~\eqn{bsa} is a good approximation for the background solution within the limits of validity of the assumptions used to derive it, and we will use it in the following sections to calculate curvature perturbations.
The analytical solution we derived around the time $\tau_0$ when the feature occurs should be accurate as long as the De sitter approximation is valid and $|\lambda^+|\ll 1$.

\section{Spectrum of curvature perturbations}
\label{curvatureperturbations}
In order to study curvature perturbations we need to expand pertubatively the action respect to the background $FRLW$ solution \cite{m,hael}. We adopt the comoving gauge in which there is no fluctuation in the scalar field, $\delta \phi=0$. The  second and third order actions are respectively
\bea
\label{s2}
 S_2&=& \int dt d^3x\left[a^3 \epsilon \dot\zeta^2-a\epsilon(\partial \zeta)^2 \right],
\eea
\bea\label{s3}
 S_3=  \int dt d^3x\left[a^3 \epsilon^2 \zeta \dot\zeta^2+ a\epsilon^2 \zeta(\partial \zeta)^2- 2a\epsilon \dot \zeta (\partial\zeta)(\partial \chi)+
 \frac{a^3\epsilon}{2}\dot \eta \zeta^2\dot \zeta \right. \\ \nonumber
 \left. + \frac{\epsilon}{2a}(\partial \zeta)(\partial \chi)\partial^2\chi + \frac{\epsilon}{4a}(\partial^2 \zeta) (\partial \chi)^2+ f(\zeta) \frac{\delta L}{\delta \zeta}\bigg|_1 \right],
\eea
where
\begin{eqnarray}
  \frac{\delta L}{\delta \zeta}\bigg|_1&=& 2a\left(\frac{d\partial^2\chi}{dt}+H \partial^2\chi-\epsilon\partial^2\zeta \right),\\
 f(\zeta) &=& \frac{\eta}{4}\zeta + \mbox{ terms with derivatives on } \zeta,
\end{eqnarray}
and $\delta L/\delta \zeta|_1$ is the variation of the quadratic action with respect to $\zeta$ \cite{m}. The Lagrange equations for the second order action give the equation for the curvature perturbations $\zeta$ 
\begin{equation}
 \frac{\partial}{\partial t}\left(a^3\epsilon \frac{\partial \zeta}{\partial t}\right)- a\epsilon\delta^{ij} \frac{\partial^2 \zeta}{\partial x^i\partial x^j}=0.
\end{equation}
The Fourier transform of the above equation,  using conformal time, gives
\begin{equation}\label{cpe}
  \zeta''_k + 2 \frac{z'}{z} \zeta'_k + k^2 \zeta_k = 0,
\end{equation}
	where $z\equiv a\sqrt{2 \epsilon}$ and $k$ is the comoving wave number. It is convenient to define the variable \cite{aer}
\begin{equation}
 u_k(\tau) \equiv z(\tau) \zeta(\tau,k),
\end{equation}
in terms of which eq.~\eqn{cpe} takes the form
\begin{equation}\label{cpe2}
 u_k''+ \left( k^2 - \frac{z''}{z} \right) u_k =0.
\end{equation}

\begin{figure}
 \begin{minipage}{.45\textwidth}
  \includegraphics[scale=0.6]{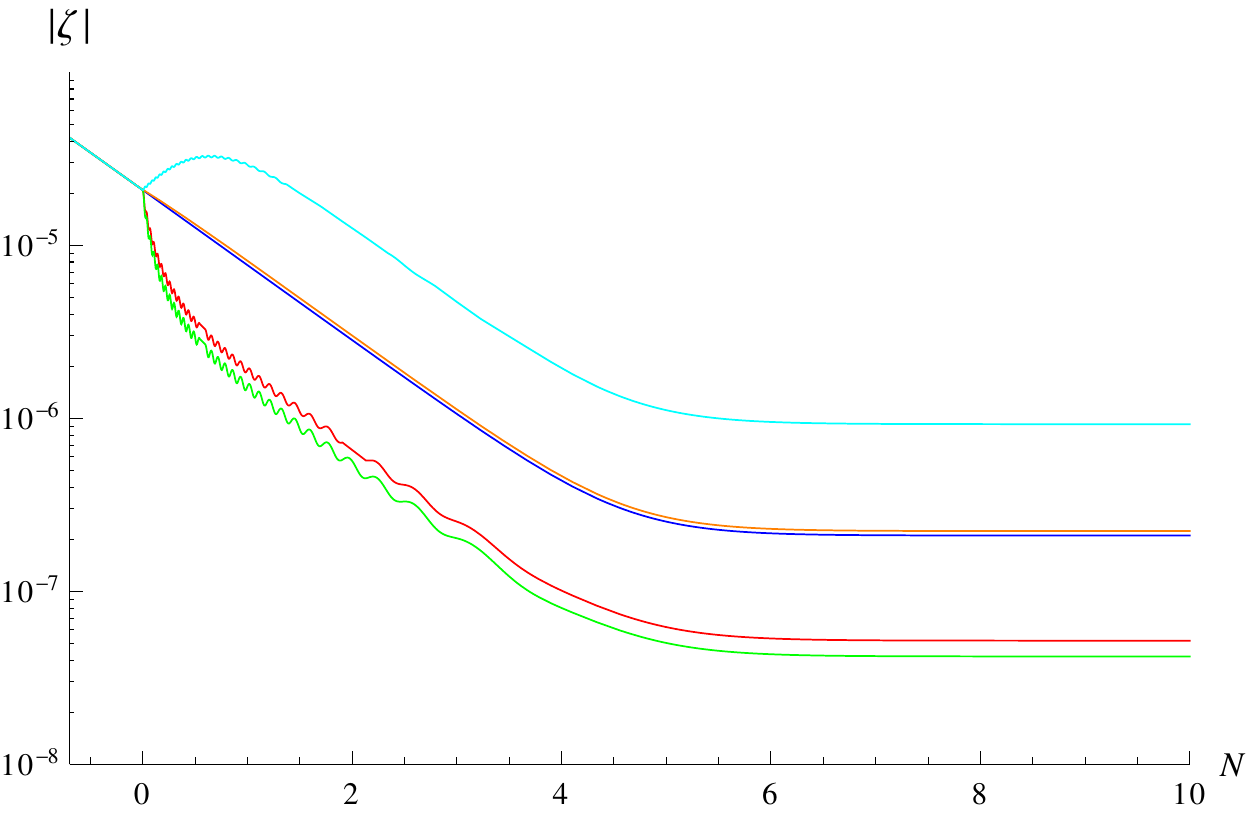}
  \end{minipage}
 \begin{minipage}{.45\textwidth}
  \includegraphics[scale=0.6]{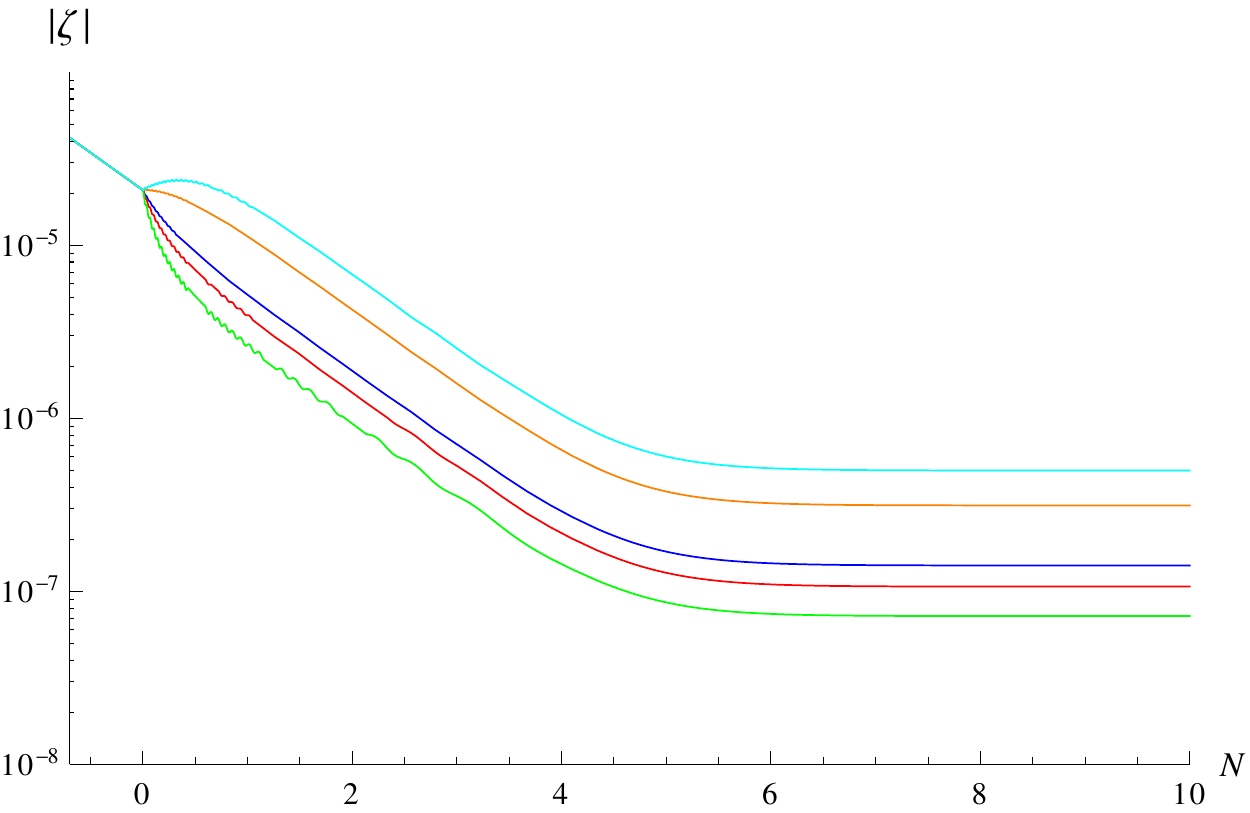}
 \end{minipage}
 \caption{The numerically computed $|\zeta_k|$ is plotted as a function of the number of $e$-folds $N$ after the time of the feature.  The plots on the left are for $\lambda=3.9\times10^{-19}$ and $n=2/3$ (blue), $n=3$ (red), and $n=4$ (green) and for $\lambda=-7\times10^{-20}$ and $n=3$ (orange) and $n=4$ (cyan). The plots on the right are for $n=3$ and $\lambda=6.0\times10^{-19}$ (blue), $\lambda=1.2\times10^{-18}$ (red), $\lambda=2.4\times10^{-18}$ (green), $\lambda=-4\times10^{-19}$ (orange), and $\lambda=-7\times10^{-19}$ (cyan). All plots are for short scale modes with $k=100k_0$ which is sub-horizon when the feature occurs.}
\label{Refn123ak0}
\end{figure}
\begin{figure}
 \begin{minipage}{.45\textwidth}
 \includegraphics[scale=0.6]{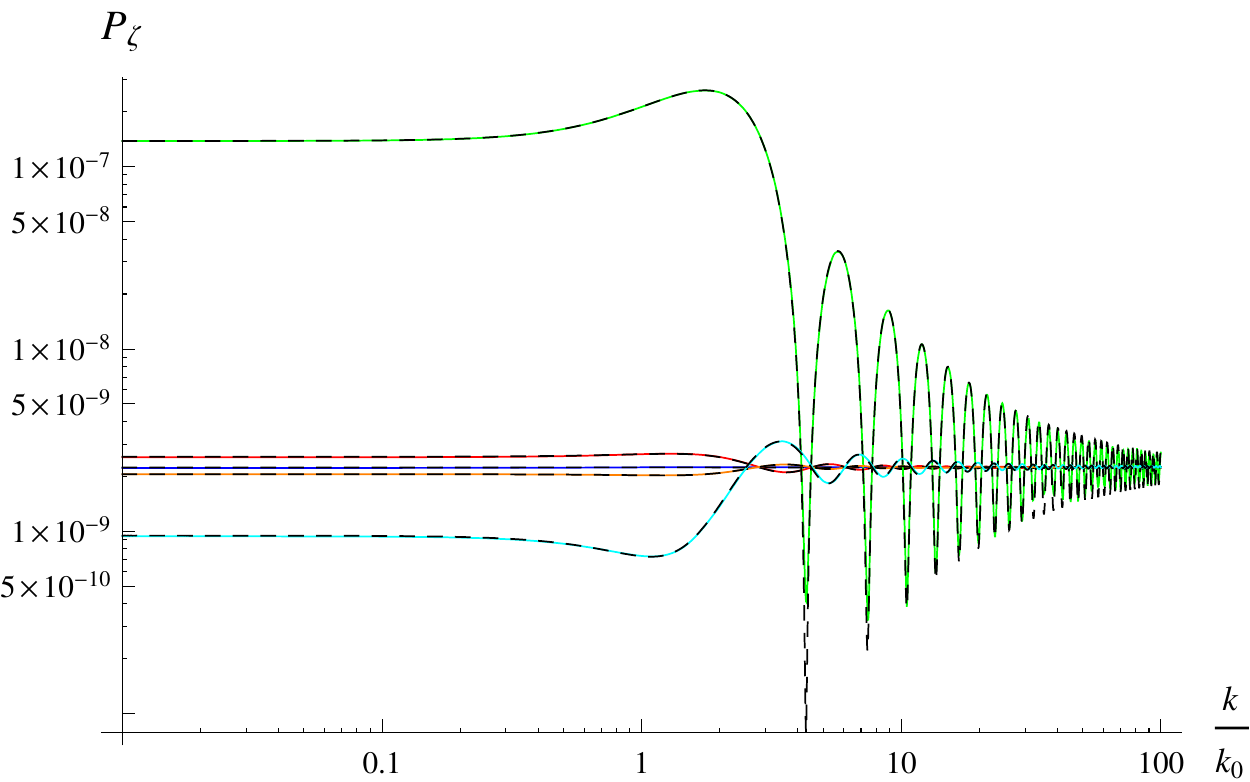}
  \end{minipage}
 \begin{minipage}{.45\textwidth}
  \includegraphics[scale=0.6]{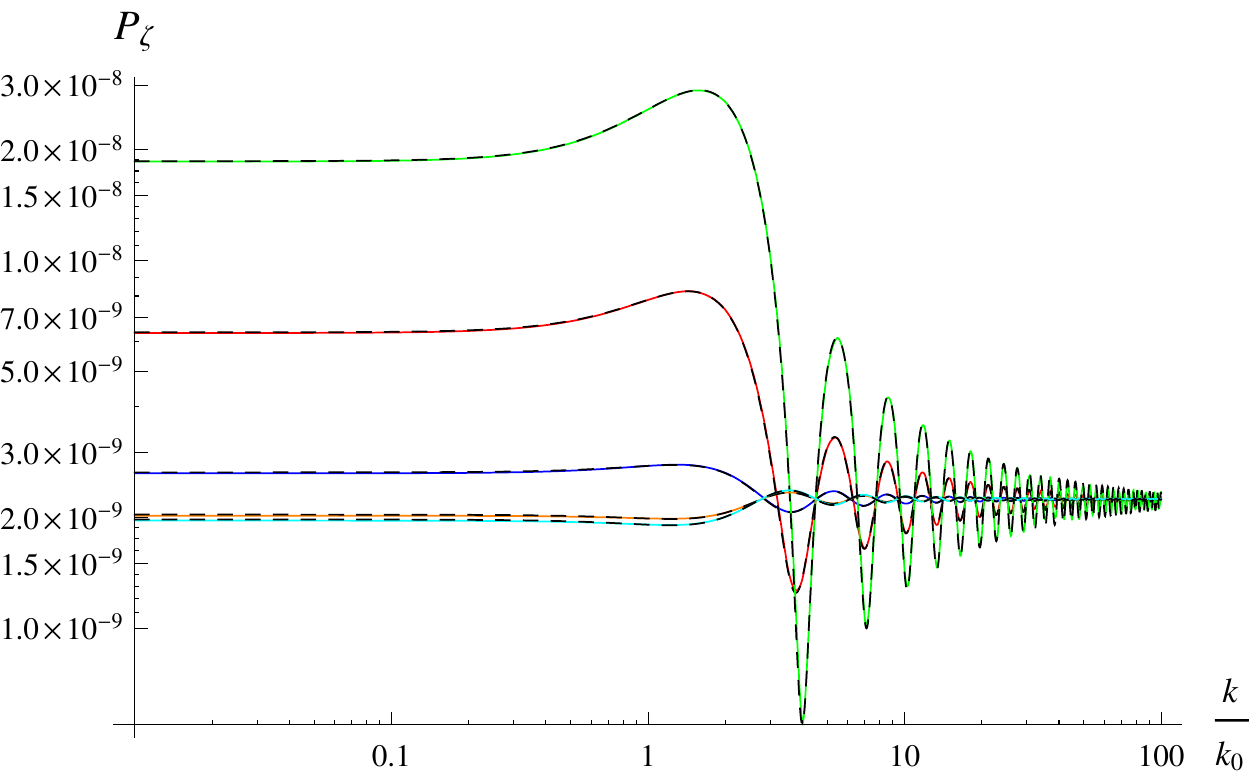}
 \end{minipage}
 \caption{The power spectrum of primordial curvature perturbations $P_{\zeta}$ is plotted for different types of features. The plots on the left are for $\lambda=8\times10^{-20}$ and $n=2/3$ (blue), $n=3$ (red), and $n=4$ (green) and for $\lambda=-5\times10^{-20}$ and $n=3$ (orange) and $n=4$ (cyan). For  right plots $n$ is constant, $n=3$, and $\lambda=1.0\times10^{-19}$ (blue), $\lambda=5\times10^{-19}$ (red), $\lambda=8\times10^{-19}$ (green), $\lambda=-8\times10^{-20}$ (orange), and $\lambda=-6\times10^{-20}$ (cyan). The dashed lines are the analytical approximations. 
%In order to satisfy the Planck normalization at small scales we set $m^2= 3H^3/(2\pi \sqrt{A_s} \phi_b^+)-n \lambda (\phi_b^+)^{n-2}$
}
\label{LogPn123}
\end{figure}

As it can be seen in fig.~(\ref{Refn123ak0}) small scales modes, which are sub-horizon at time $\tau_0$, are affected by the feature.  Modes that had  left the horizon at that time are unaffected, since they were already frozen. 
In fig.~(\ref{LogPn123}) the power spectrum of primordial curvature perturbations $P_{\zeta}$ is plotted for different types of features.
% and for $\lambda=-7\times10^{-20}$ and $n=3$ (orange) and $n=4$ (cyan)
%, $\lambda=-4\times10^{-19}$ (orange), and $\lambda=-7\times10^{-19}$ (cyan)

%\section{Numerical calculation of the spectrum}

\begin{figure}
  \includegraphics[scale=1]{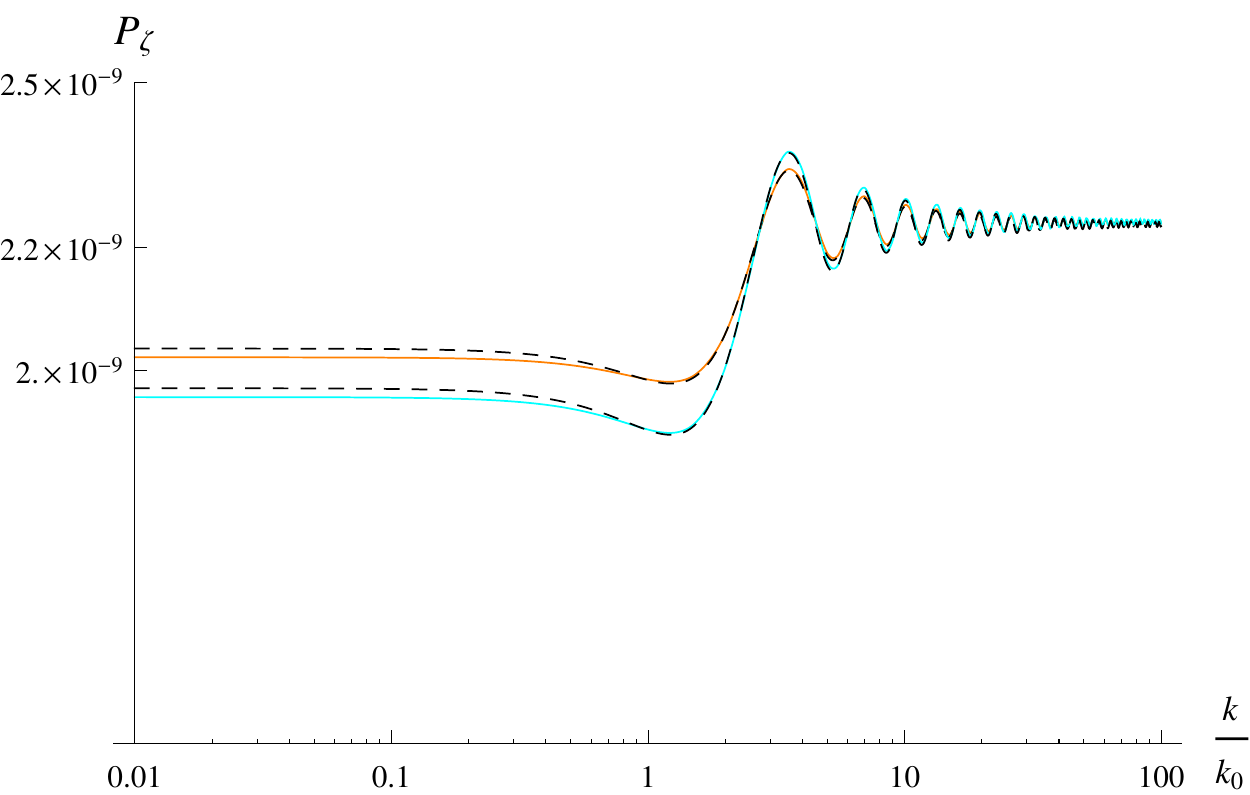}
 \caption{The power spectrum of primordial curvature perturbations $P_{\zeta}$ is plotted for  $n=3$, %$m^2= 3H^3/(2\pi \sqrt{A_s} \phi_b^+)-n \lambda (\phi_b^+)^{n-2}$, 
 $\lambda=-8\times10^{-20}$ (orange) and $\lambda=-6\times10^{-20}$ (cyan). The dashed lines are the analytical approximations. These models are able to account for the  observed large scale suppression. }
\label{PPS}
\end{figure}

\section{Analytical approximation for  curvature perturbations}
At the time of the feature there is a discontinuity in $\phi''$ which implies that $z''$ contains a Dirac delta function \cite{pp}. 
We can evaluate the discontinuity in $z''/z$ by integrating the Dirac delta function around the feature time  
\begin{equation}\label{d0}
  D_0 \equiv \lim_{\delta \to 0} \int_{\tau_0-\delta}^{\tau_0+\delta}\frac{z''}{z}d\tau = \frac{1}{\phi_0'} \left[ \phi''_{a0}-\phi''_{b0} \right]
  =-n \lambda a(\tau_0)^2 \frac{\phi_0^{n-1}}{\phi_0'}.
\end{equation}
Before the feature we assume the  Bunch-Davies vacuum \cite{Bunch:1978yq}
\begin{equation}
  v(\tau,k)=\frac{e^{-\mathrm{i} k\tau}}{\sqrt{2k}}\left(1-\frac{\mathrm{i}}{k\tau}\right).
\end{equation}

\begin{figure}
 \begin{minipage}{.45\textwidth}
  \includegraphics[scale=0.6]{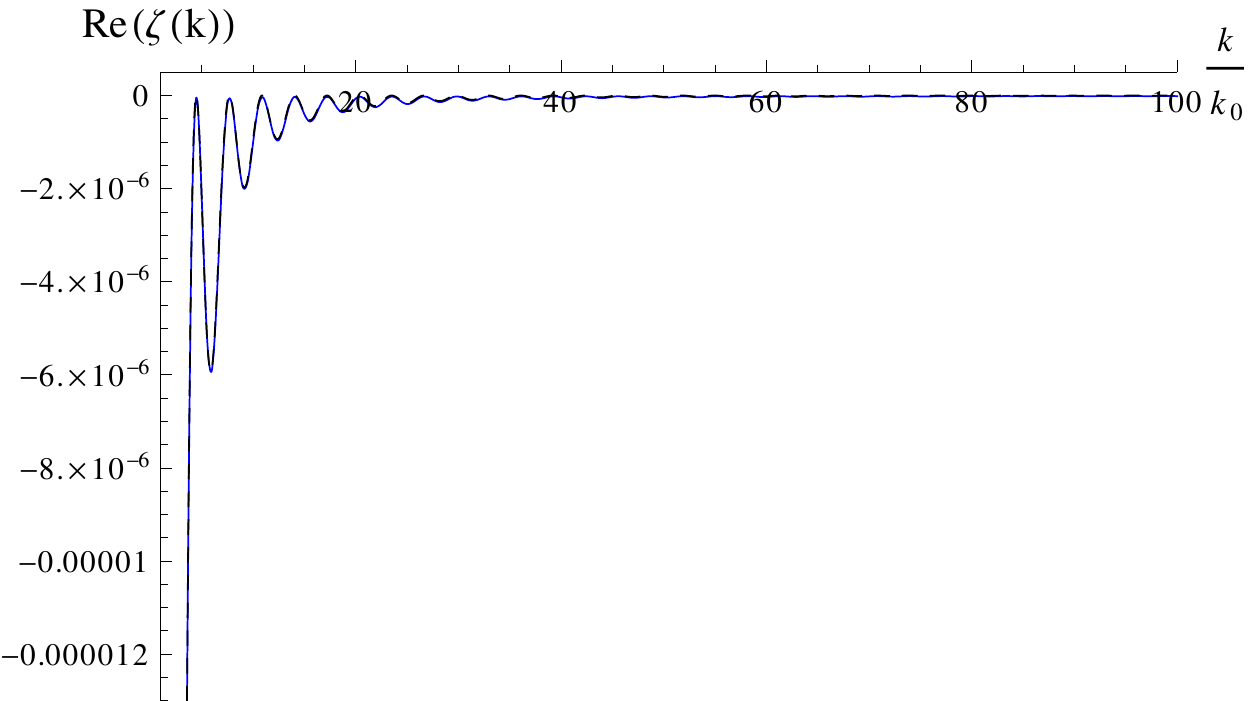}
  \end{minipage}
 \begin{minipage}{.45\textwidth}
  \includegraphics[scale=0.6]{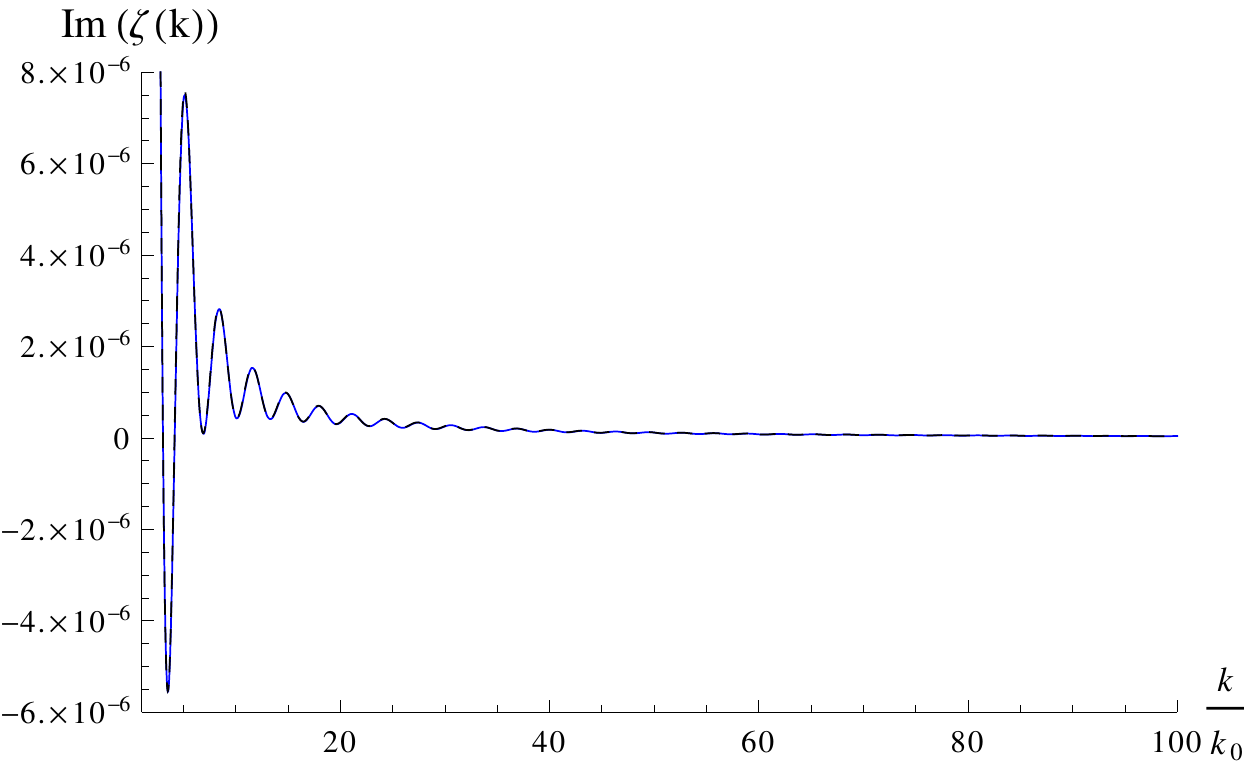}
 \end{minipage}
 \caption{Numerically  (blue)  and analytically  (black lines) computed  small scale modes evaluated at $10e$-folds after the feature are plotted as functions of the scale. On the left it is plotted the real part, on the right the imaginary part. The parameters used for the feature are $n=4$ and $\lambda=3.9\times10^{-19}$.}
\label{Rteka}
\end{figure}

\begin{figure}
 \begin{minipage}{.45\textwidth}
  \includegraphics[scale=0.6]{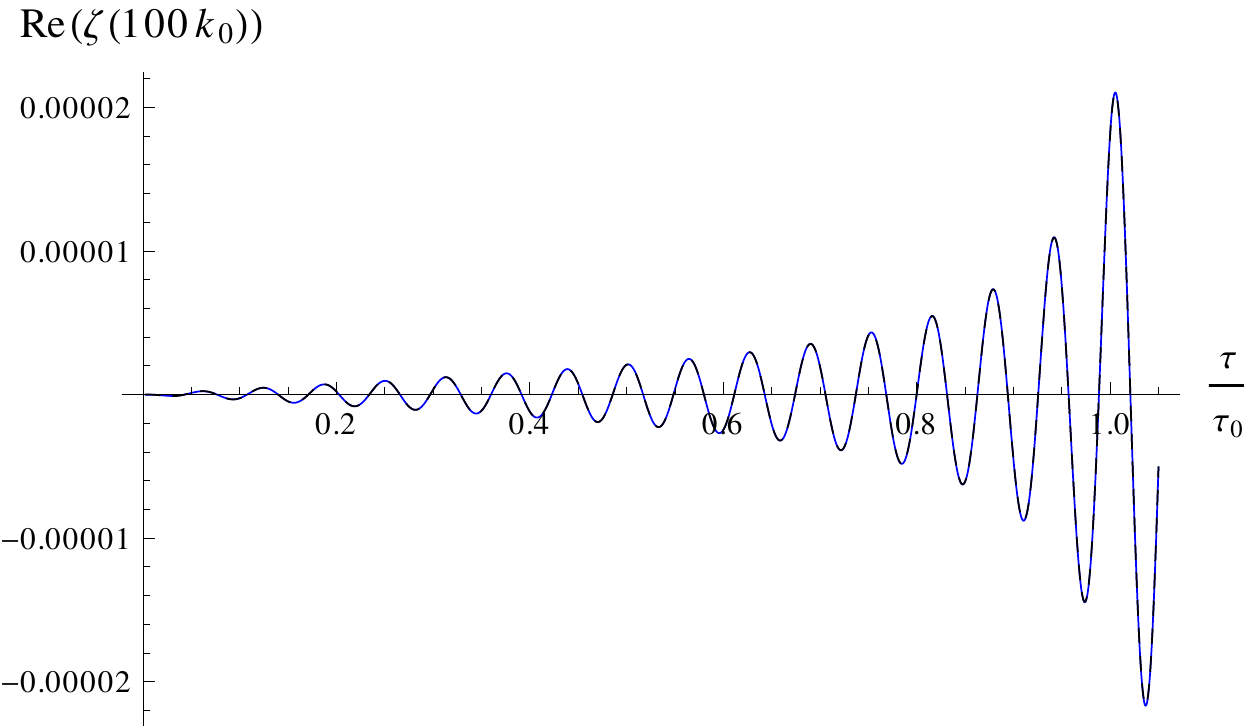}
  \end{minipage}
 \begin{minipage}{.45\textwidth}
  \includegraphics[scale=0.6]{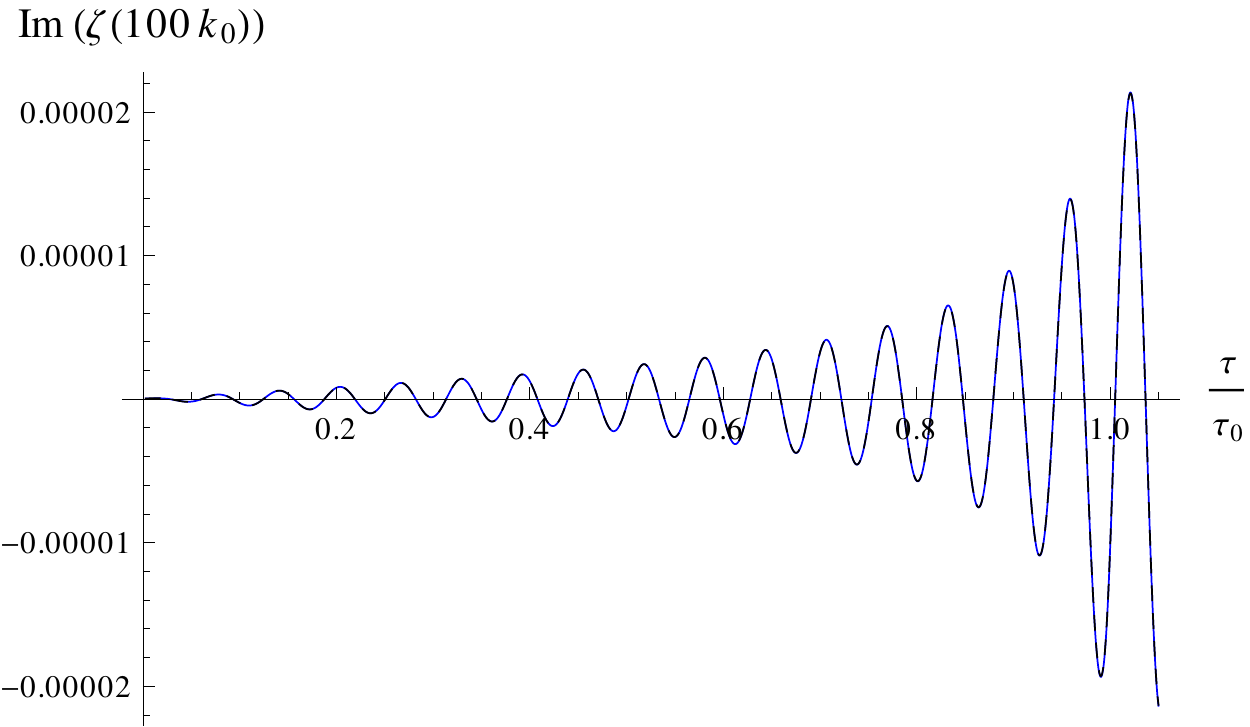}
 \end{minipage}
 \caption{Comparison of the evolution of the real (left) and imaginary (right) parts of the mode function  for $k=100k_0$. The result of numerical calculations is plotted in blue, while the analytical approximation is plotted in black. The parameters used for the feature are $n=4$ and $\lambda=3.9\times10^{-19}$.}
\label{Rka100}
\end{figure}

The curvature perturbations modes after the feature are approximated as a linear combination of the positive and negative frequency modes before the feature \cite{starobinsky,Starobinsky:1998mj} according to
\begin{equation}\label{r}
  \zeta(\tau,k)=\frac{1}{ a(\tau)\sqrt{2\epsilon(\tau)} }\left[\alpha(k)v(\tau,k)+\beta(k)v^*(\tau,k)\right] \,,
\end{equation}
where
\begin{equation}\label{alphabeta}
  \alpha(k)=1+ \mathrm{i} D_0 |v(\tau_0,k)|^2 \: \mbox{ and } \: \beta(k)= -\mathrm{i} D_0 v(\tau_0,k)^2
	\end{equation}
are the Bogoliubov coefficients, and $\tau_k=-1/k$ is the horizon crossing time for the mode $k$. The coefficients $\alpha,\beta$ are determined by imposing the continuity of the modes and their derivative at the feature time \cite{pp}. In figs.~(\ref{Rteka}-\ref{Rka100}) we show the comparison between the numerical results for the mode function and the analytical approximation for small scales. The parameters used for the feature are $n=4$ and $\lambda=3.9\times10^{-19}$. In fig.~(\ref{Rteka}) we show  the real and imaginary part of the dependence of the mode functions on the scale and for a particular time, namely, after ten $e$-folds after the feature. While in fig.~(\ref{Rka100}) we show the evolution of the real and imaginary parts of the mode
function at a particular scale $100k_0$. 
%The comparison between the numerical and the analytical approximation for other values of $n$ and $\lambda$ are also in good agreement.

\section{Analytical approximation for the spectrum}
The two-point function is
\begin{equation}
 \Braket{ \zeta(\vec{k}_1, t) \zeta(\vec{k}_2, t) } \equiv (2\pi)^3 \frac{2\pi^2}{k^3} P_{\zeta}(k) \delta^{(3)}(\vec{k}_1+\vec{k}_2) \, ,
\end{equation}
where the power spectrum of curvature perturbations is defined as 
\begin{equation}\label{ps2}
  P_{\zeta}(k) \equiv \frac{k^3}{2\pi^2}|\zeta_k(\tau_e)|^2,
\end{equation}
where $\tau_e$ is the time at which inflation ends. After substituting in the above definition the analytical approximations obtained in the previous sections  we get
\bea\label{mps}
P_{\zeta}(k)=\frac{ H^2}{8\pi^2 \epsilon(\tau_e)} \left\{1 + \frac{D_0}{k} \left[\left(\frac{k_0^2}{k^2}-1\right)\sin\left(\frac{2 k}{k_0}\right)-\frac{2k_0}{k} \cos\left(\frac{2 k}{k_0}\right) \right] \right. \nonumber \\ \\ 
	\left. +\frac{D_0^2}{2 k^2} \left[ 1+ \frac{2k_0^2}{k^2} +\frac{k_0^4}{k^4}+\left(1-\frac{k_0^4}{k^4}\right)\cos\left(\frac{2 k}{k_0}\right)-\frac{2k_0}{k}\left(1+\frac{k_0^2}{k^2}\right)\sin\left(\frac{2 k}{k_0}\right)\right] \right\} \, . \nonumber 
\eea
This generalizes the result obtained in \cite{starobinsky} for a potential with a discontinuous first derivative to the more general case considered in this paper, corresponding to  eq.~(36). 
In fig.~(\ref{LogPn123}) we compare the analytical expression for the power spectrum given by eq.~\eqn{mps} with the numerical results obtained by integrating numerically both the background and the perturbations equations.
The analytical result is quite accurate and it improves the results obtained in \cite{aer} because  we use the analytical approximation for the perturbations modes also for modes which were superhorizon slightly before $\tau_0$, improving substantially the agreement with numerical results. This is due to the well known fact that modes are not completely frozen at $\tau_k=-1/k$, but keep evolving for few e-folds after, so that also scales slightly greater than $k_0$ are mildly affected by the features.

It should be noted that the formula (\ref{mps}) we obtained is not depending on the slow-roll approximation, since it is derived directly from the definition of the power spectrum in eq.~(\ref{ps2}), and this explains why it is in such a good agreement with fully numerical calculations despite the temporary violation of slow-roll regime produced by the features.
The $\epsilon$  in the denominator of eq.~(\ref{mps}) comes in fact from the analytical solution for the perturbation modes, which is not based on any slow-roll expansion because it comes from $z\equiv a\sqrt{2 \epsilon}$ in eqs.~(\ref{cpe},\ref{cpe2}) which are valid at any order in slow-roll.

As can be seen in fig.~(\ref{LogPn123}) negative values of $\lambda$ correspond to a suppression of the spectrum on large and intermediate scales, while positive values produce a suppression on small scales.
While the analysis of observational data goes beyond the scope of this paper, we can see in fig.~(\ref{LogPn123}) that an appropriate choice of parameters gives spectra in good qualitative agreement with the features recently found when parameterizing the free primordial power spectrum with
a piecewise cubic Hermite interpolating polynomial \cite{Gariazzo:2014dla}.

\section{Calculation of the bispectrum}
The Fourier transform of the three-point correlation function  \cite{pxxii, pxxiv}, also known as the bispectrum $B_{\zeta}$ is given by
\begin{equation}
 \Braket{ \zeta(\vec{k}_1, t) \zeta(\vec{k}_2, t)  \zeta(\vec{k}_3, t) }= (2\pi)^3 B_{\zeta}(k_1,k_2,k_3) \delta^{(3)}(\vec{k}_1+\vec{k}_2,\vec{k}_3),
\end{equation}
and it should vanish if the curvature perturbations are Gaussian \cite{xc,a2}. Therefore, deviations from non Gaussianity can be determined by measuring the implications of a non vanishing $B_{\zeta}$ on the CMB radiation. Following the procedure in Ref. \cite{aer}, after a field redefinition, the third order action can be written as
\begin{equation}
 S_3=   \int dt d^3x\left[-a^3 \epsilon \eta \zeta \dot\zeta^2-\frac{1}{2}a \epsilon\eta \zeta \partial^2 \zeta \right].
\end{equation}
From this action the interaction Hamiltonian can be written in terms of conformal time as	
\bea
H_{int}(\tau)=  \int d^3x \, \epsilon \eta a \left[ \zeta \zeta'{}^2 + \frac{1}{2} \zeta^2 \partial^2 \zeta \right].
\eea
The 3-point correlation function is given by \cite{m,xc}
\bea
\Braket{\Omega| \zeta(\tau_e,\vec{k}_1) \zeta(\tau_e,\vec{k}_2)  \zeta(\tau_e,\vec{k}_3)|\Omega }= -\mathrm{i} \int_{-\infty}^{\tau_e} \Braket{0|\left[ \zeta(\tau_e,\vec{k}_1) \zeta(\tau_e,\vec{k}_2)  \zeta(\tau_e,\vec{k}_3), H_{int}\right]|0 },
\eea
and after substitution the expression for the bispectrum $B_{\zeta}$ \cite{inin,xc} is
\bea \label{b}
  B_{\zeta}(k_1,k_2,k_3)= 2 \Im\Bigl[ \zeta(\tau_e,k_1) \zeta(\tau_e,k_2)\zeta(\tau_e,k_3) \int^{\tau_e}_{\tau_0} d\tau \eta \epsilon a^2 
 \biggl( 2\zeta^*(\tau,k_1) \zeta'{}^*(\tau,k_2)\zeta'{}^*(\tau,k_3)  \\ \nonumber 
 - k^2_1 \zeta^*(\tau,k_1)\zeta^*(\tau,k_2)\zeta^*(\tau,k_3) \biggr)   
+ \mbox{ two permutations of } k_1, k_2,\mbox{ and } k_3  \Bigr]\, ,
\eea
where $\Im$ is the imaginary part and we evaluate the integral from $\tau_0$ to $\tau_e$, where $\tau_e$ is some time sufficiently after Hubble crossing horizon, when the modes are frozen \cite{a1,a2,a3, bingo,numerical2013}. 

It is common to study  non-Gaussianity using the  parameter $f_{NL}$ defined by 
\bea\label{fNL}
\frac{6}{5} f_{NL}(k_1,k_2,k_3)\equiv \frac{B_{\zeta}}{\mathbf{P}_{\zeta}(k_1)\mathbf{P}_{\zeta}(k_2)+\mathbf{P}_{\zeta}(k_1)\mathbf{P}_{\zeta}(k_3)+\mathbf{P}_{\zeta}(k_2)\mathbf{P}_{\zeta}(k_3)} \, ,
\eea
where
\begin{equation}
\mathbf{P}_{\zeta} \equiv \frac{2\pi^2}{k^3} P_{\zeta} \, ,
\end{equation}
Replacing $\mathbf{P}_{\zeta}$ in eq.(\ref{fNL}) we obtain  $f_{NL}$ in terms of our dimensionless definition of the spectrum $P_{\zeta}(k)$
\bea
f_{NL}(k_1,k_2,k_3)= \frac{10}{3}\frac{(k_1 k_2 k_3)^3}{(2\pi)^4} \frac{B_{\zeta}}{P_{\zeta}(k_1)P_{\zeta}(k_2)k^3_3+P_{\zeta}(k_1)P_{\zeta}(k_3)k^3_2+P_{\zeta}(k_2)P_{\zeta}(k_3)k^3_1} \, .
\eea
In this paper we will study non-Gaussianity using a different quantity defined as
\begin{equation}\label{FNL}
  F_{NL}(k_1,k_2,k_3;k_*)\equiv \frac{10}{3(2\pi)^4}\frac{(k_1 k_2 k_3)^3}{k_1^3+k_2^3+k_3^3}\frac{B_{\zeta}(k_1,k_2,k_3)}{P_{\zeta}^2(k_*)} \, ,
\end{equation}
where $k_*$ is a  pivot scale at which the power spectrum is evaluated which corresponds approximately to the scale of normalization of the spectrum, i.e. $P_{\zeta}(k_*)\approx 2.2\times 10^{-9}$.
Our definition of $F_{NL}$ reduces to $f_{NL}$ in the equilateral limit if the spectrum is approximately scale invariant, but in general $f_{NL}$ and $F_{NL}$ are different, and for example in the squeezed limit they are not the same. For this reason they cannot be compared directly but $F_{NL}$ still provides useful information about the non-Gaussian behavior of $B_{\zeta}$.

\begin{figure}
 \begin{minipage}{.45\textwidth}
  \includegraphics[scale=0.6]{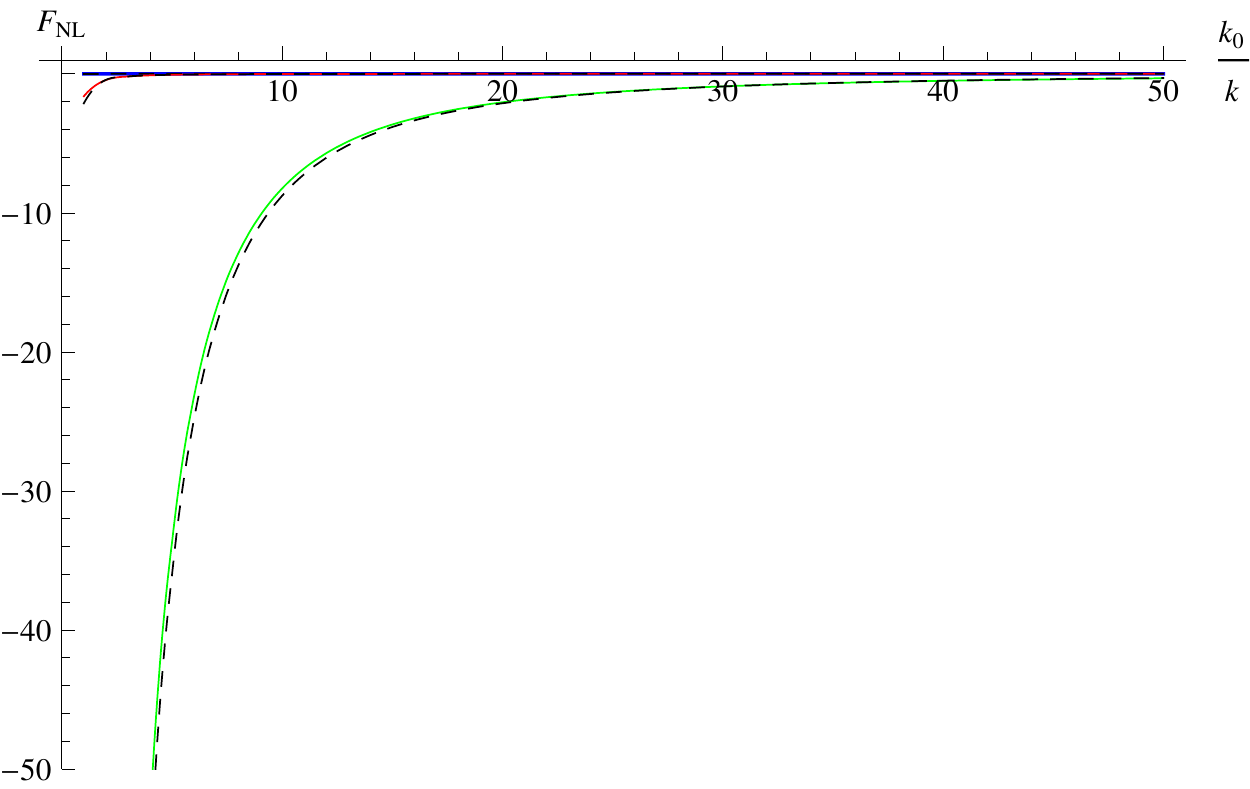}
  \end{minipage}
 \begin{minipage}{.45\textwidth}
  \includegraphics[scale=0.6]{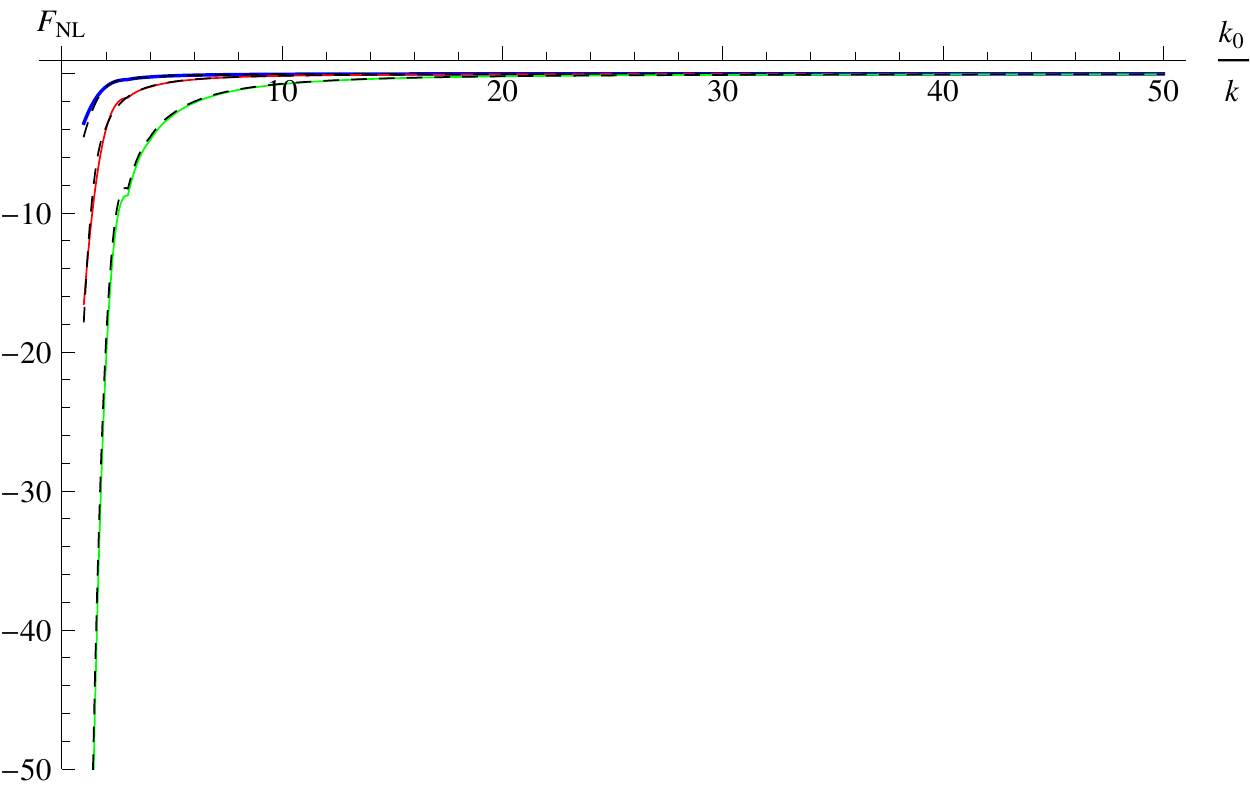}
 \end{minipage}
 \caption{The squeezed limit of the numerically computed bispectrum  $F_{NL}(k_0/500,k,k)$ in plotted for a large scale $k_0/500$. On the left we keep $\lambda$ constant, $\lambda=3.9\times10^{-19}$, while $n=2/3$ (blue), $n=3$ (red), and $n=4$ (green). On the right we keep $n$ constant, $n=3$, while $\lambda=6.0\times10^{-19}$ (blue), $\lambda=1.2\times10^{-18}$ (red), and $\lambda=2.4\times10^{-18}$ (green). The dashed black lines correspond to the analytical approximation.}
\label{FNLfnlslb}
\end{figure}

\begin{figure}
 \begin{minipage}{.45\textwidth}
  \includegraphics[scale=0.6]{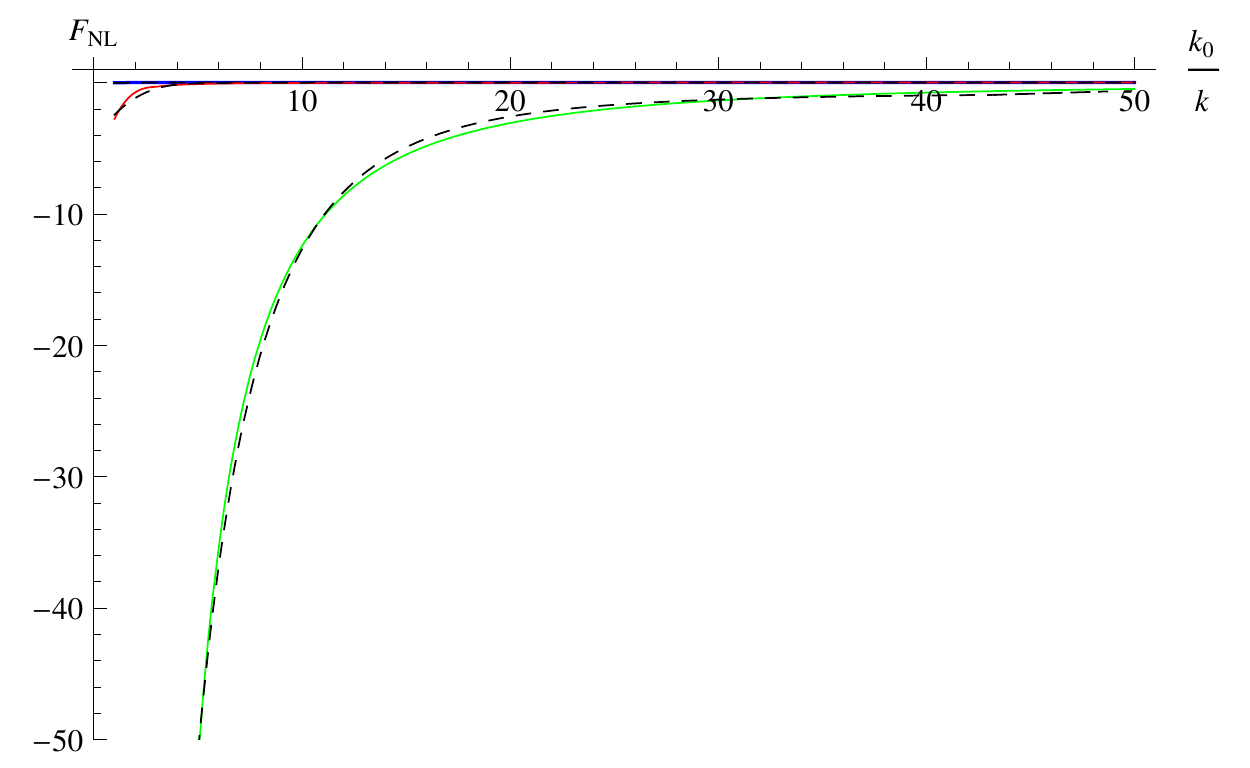}
  \end{minipage}
 \begin{minipage}{.45\textwidth}
  \includegraphics[scale=0.6]{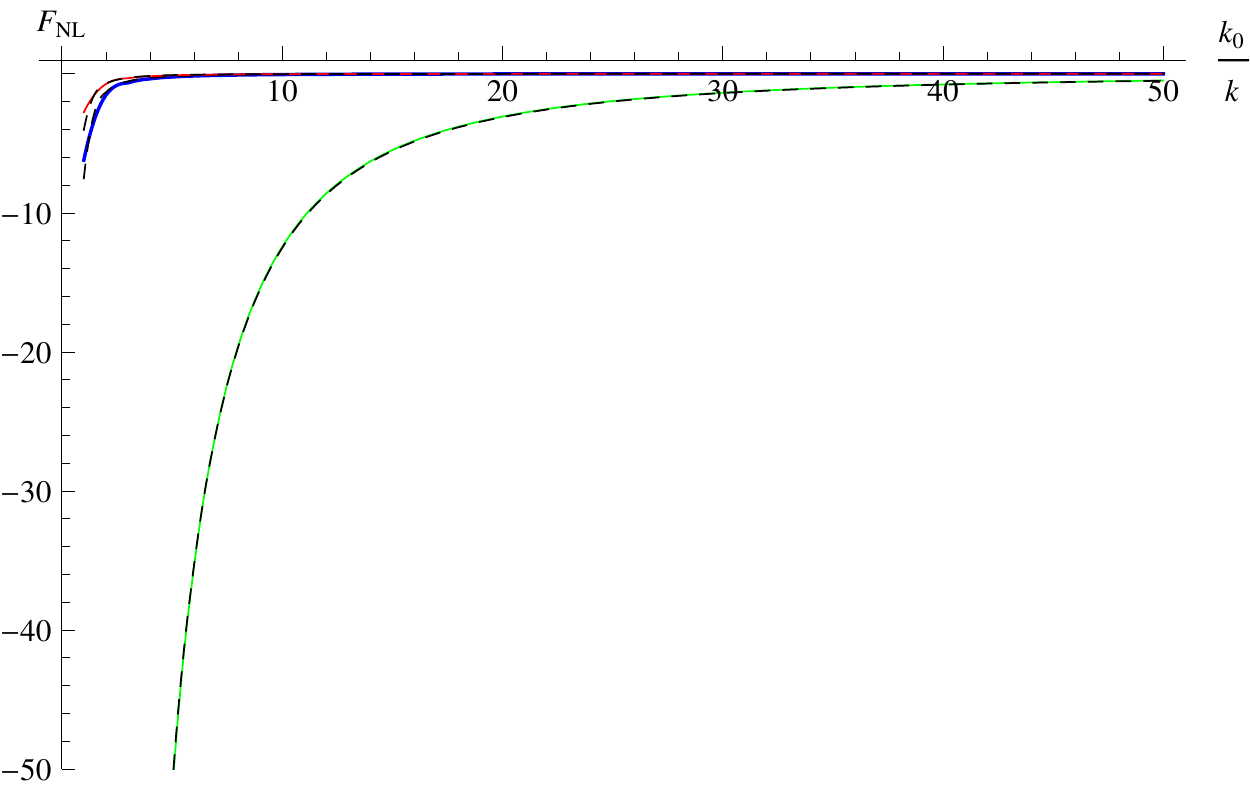}
 \end{minipage}
 \caption{The equilateral limit of the numerically computed bispectrum  $F_{NL}(k,k,k)$ in plotted for large scales. On the left we keep $\lambda$ constant, $\lambda=3.9\times10^{-19}$, while $n=2/3$ (blue), $n=3$ (red), and $n=4$ (green). On the right we keep $n$ constant, $n=3$, while $\lambda=6.0\times10^{-19}$ (blue), $\lambda=1.2\times10^{-18}$ (red), and $\lambda=2.4\times10^{-18}$ (green). The dashed black lines correspond to the analytical approximation.}
\label{FNLfnlelb}
\end{figure}

\begin{figure}
 \begin{minipage}{.45\textwidth}
  \includegraphics[scale=0.6]{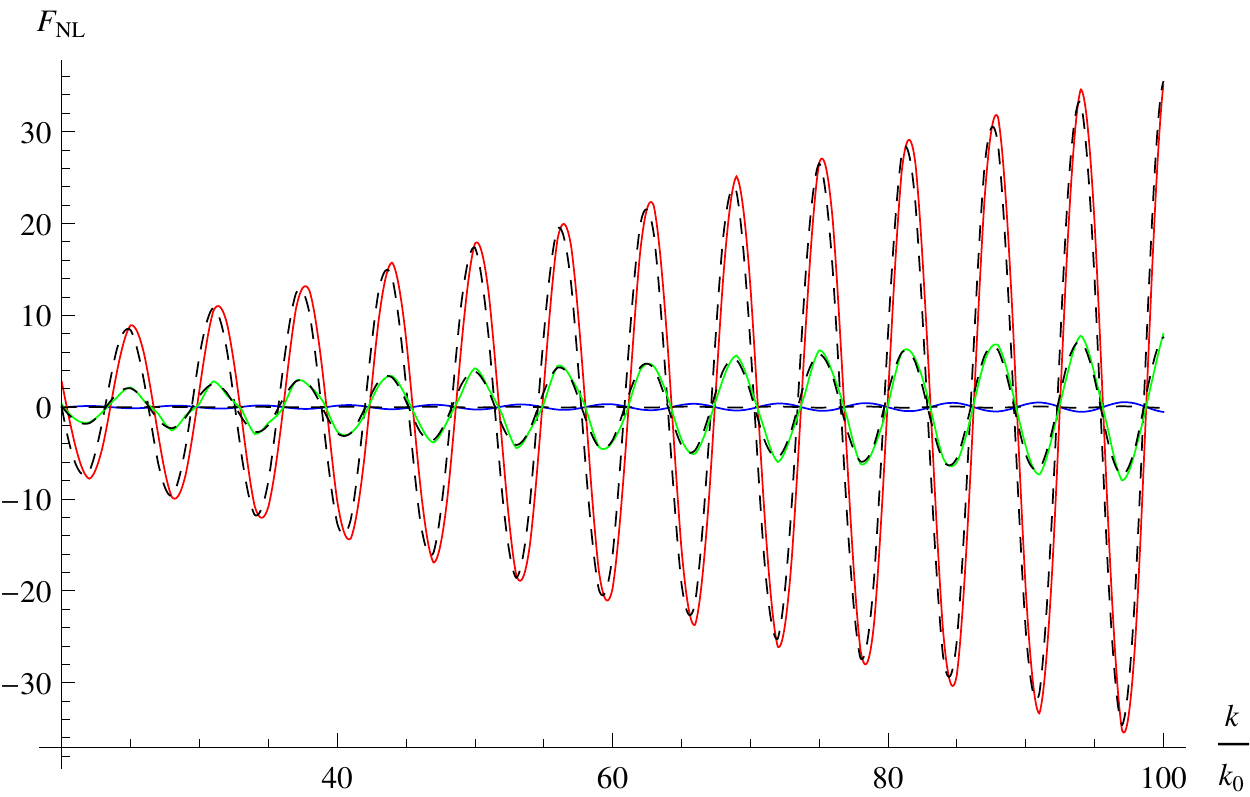}
  \end{minipage}
 \begin{minipage}{.45\textwidth}
  \includegraphics[scale=0.6]{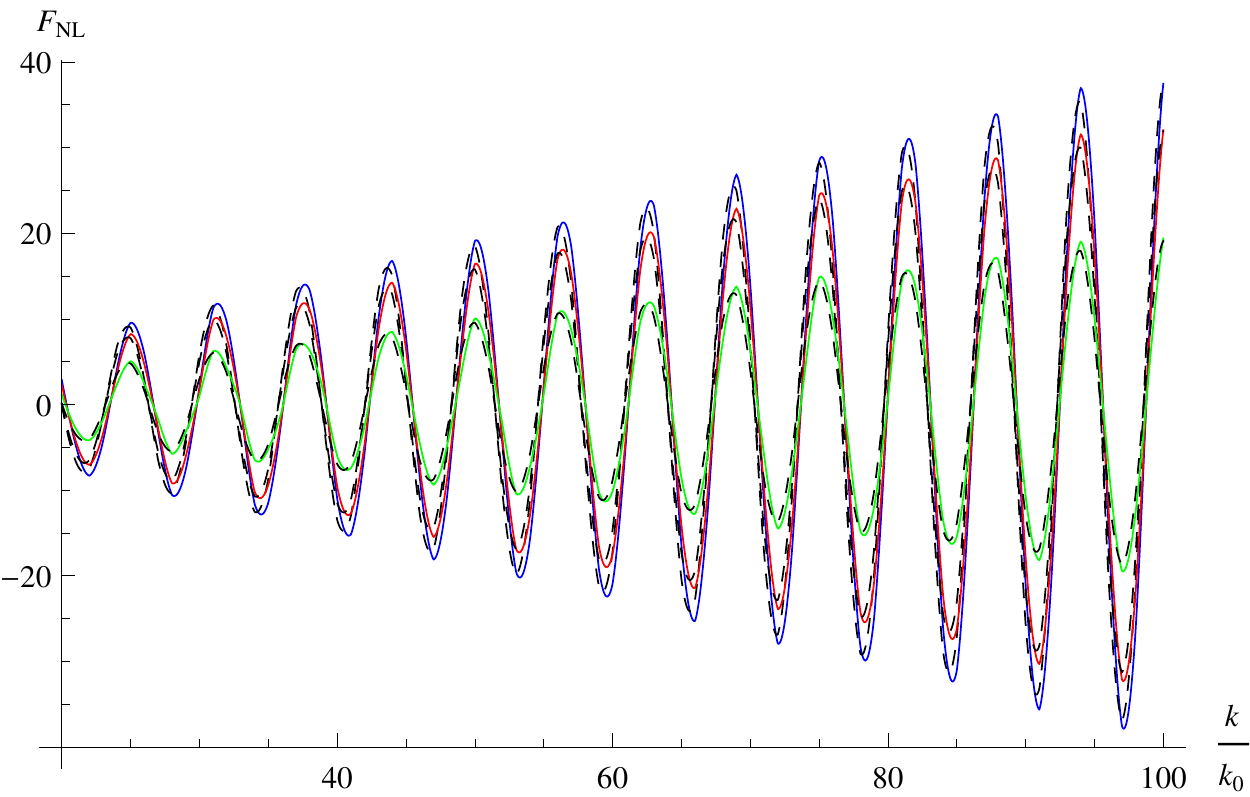}% -------------------------------------------------------
 \end{minipage}
 \caption{The squeezed limit of the numerically computed bispectrum  $F_{NL}(k,1000k_0,1000k_0)$ is plotted for a small scale $1000 k_0$. On the left $\lambda$ is constant, $\lambda=3.9\times10^{-19}$, while $n=2/3$ (blue), $n=3$ (red), and $n=4$ (green). On the right $n$ is constant, $n=3$, while $\lambda=6.0\times10^{-19}$ (blue), $\lambda=1.2\times10^{-18}$ (red), and $\lambda=2.4\times10^{-18}$ (green). The dashed black lines correspond to the analytical approximation.}
\label{FNLfnlsla}
\end{figure}

\begin{figure}
 \begin{minipage}{.45\textwidth}
  \includegraphics[scale=0.6]{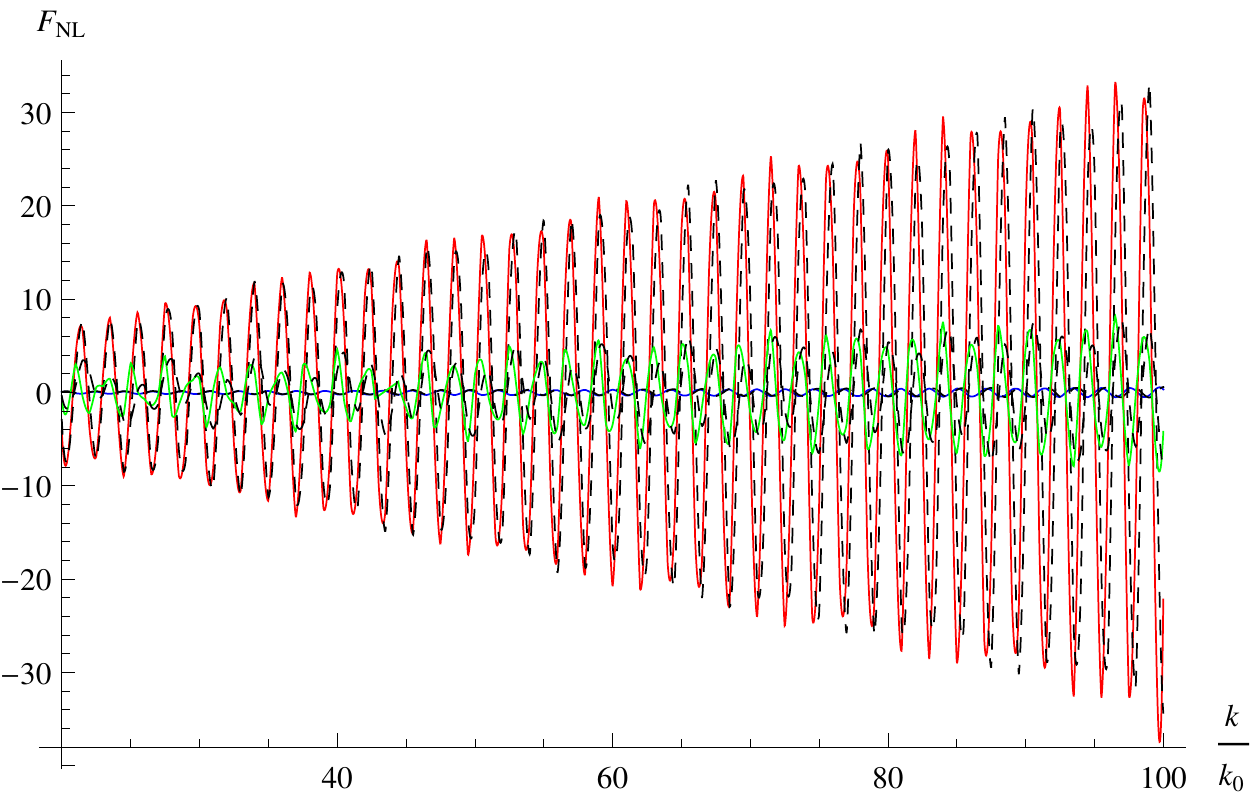}
  \end{minipage}
 \begin{minipage}{.45\textwidth}
  \includegraphics[scale=0.6]{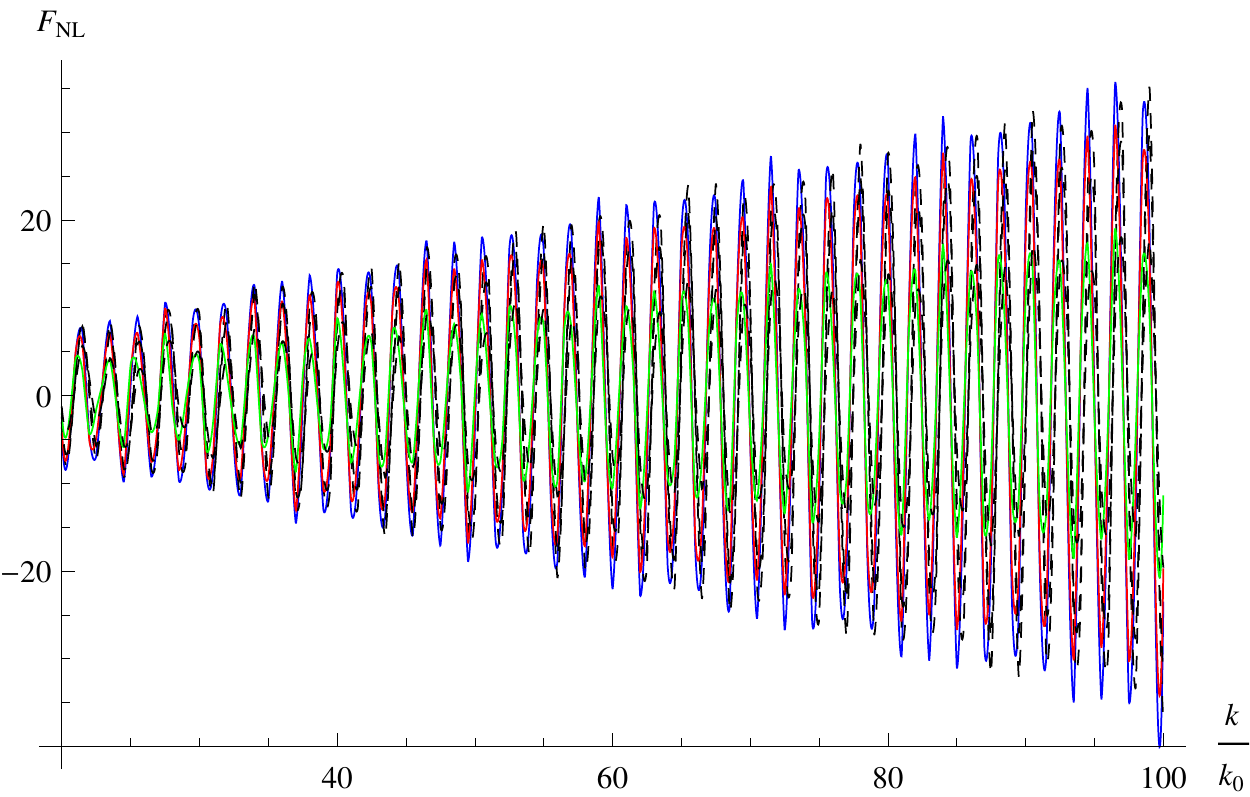}
 \end{minipage}
 \caption{The equilateral limit of the numerically computed bispectrum  $F_{NL}(k,k,k)$ is plotted for small scales. On the left $\lambda$ is constant, $\lambda=3.9\times10^{-19}$, while $n=2/3$ (blue), $n=3$ (red), and $n=4$ (green). On the right $n$ is constant, $n=3$, while $\lambda=6.0\times10^{-19}$ (blue), $\lambda=1.2\times10^{-18}$ (red), and $\lambda=2.4\times10^{-18}$ (green). The dashed black lines correspond to the analytical approximation.}
\label{FNLfnlela}
\end{figure}

In figs.~(\ref{FNLfnlslb}-\ref{FNLfnlelb}) the large scale squeezed and equilateral limit of the bispectrum are plotted for different values of the parameters $n$ and $\lambda$. The small scale squeezed and equilateral limits are shown in figs.~(\ref{FNLfnlsla}) and (\ref{FNLfnlela}), respectively. 
%In figs.~(\ref{FNLfnlslb}-\ref{FNLfnlelb}) it can be seen that in the case of $\lambda$ constant and $n=2/3$ the effect of the feature is almost null which can be expected since, as we mentioned before, the increment of the slow roll parameter $\eta$ is very small compared to the case without feature [see fig.~(\ref{Etan123})].
As shown in figs.~(\ref{FNLfnlsla}-\ref{FNLfnlela})  the bispectrum has an oscillatory behavior with an amplitude inversely proportional to the scale.

\section{Analytical approximation for the bispectrum}
In order to obtain an analytical approximation for the bispectrum we use eq.~\eqn{r} for curvature perturbations and  eq.~\eqn{slowrollapprox} for slow-roll parameters. This implies that also the 
different approximations are used in different cases as explained in more details in the following sections.
All the results presented in this section should be considered taking into account the existence of a cut-off scales beyond which the Heaviside approximation of a smooth transition is not valid as discussed in more details in \cite{aer,pp}.
We provide analytical expressions for the bispectrum for different types of features, i.e., different values of $n$ and $\lambda$, in the squeezed and equilateral limit for both large and small scales. The analytical results are shown in dashed black lines in figs.~(\ref{FNLfnlslb}-\ref{FNLfnlela}) where it can be seen that the approximations for the bispectrum  are in good agreement with the numerical results.

%The formulas we obtain are also based on pulling out of the integrals the modes after horizon crossing, since they are frozen, and %on taking the slow roll parameters as constant except in the time interval around when the feature occurs, when in fact their %variation cannot be neglected. 
\subsection{Large scales}
In the large scale limit when $k_i < k_0, \,i=1,2,3$, the curvature perturbations modes are frozen in the time interval of interest, since there is no time evolution for $\tau >\tau_0>\tau_{k_i}$. Thus in eq.~\eqn{b} all the modes functions can be evaluated at $\tau_0$ and pulled out of the integrals while the terms $\zeta'{}^*(\tau,k_i)$ can be set to zero. 
Following this approximation we get 
\bea
  F_{NL}^<(k_1,k_2,k_3) \approx \frac{-20(\prod_{i=1}^{3}k_i^3)}{3(2\pi)^4 P_{\zeta}^2(k_*)}\frac{\sum_{i=1}^{3} k_i^2}{\sum_{i=1}^{3}k_i^3} \Im \left[\prod_{i=1}^{3}\zeta(\tau_e,k_i)\zeta^*(\tau_0,k_i)  \right] \int^{\tau_e}_{\tau_0} d\tau \eta\epsilon a^2   \\ \nonumber 
  \approx \frac{-20(\prod_{i=1}^{3}k_i^3)}{3(2\pi)^4 P_{\zeta}^2(k_*)}\frac{\sum_{i=1}^{3} k_i^2}{\sum_{i=1}^{3}k_i^3} \frac{\phi_a^{+2} (\lambda^+)^3 a(\tau_e)}{H} \Im \left[ \prod_{i=1}^{3}\zeta(\tau_e,k_i)\zeta^*(\tau_0,k_i) \right] \, ,
\eea
where we have used the approximations for the slow-roll parameters in eq.~\eqn{slowrollapprox} in the integration. Now we use the analytical approximations for the perturbation to obtain at large scales
\bea\label{fnlsl}
F_{NL}^<(k_1,k_2,k_3) \approx -\frac{5}{6}\frac{H^5}{(2\pi)^4P_{\zeta}^2(k_*)} \frac{\phi_a^{+2}a(\tau_e)}{(\lambda^+)^3 (\phi_b^+)^6} \frac{\sum_{i=1}^{3} k_i^2}{\sum_{i=1}^{3}k_i^3} \Im \left[\prod_{i=1}^{3}(\tau_e k_i-\mathrm{i}) (k_i \tau_0 + \mathrm{i}) e^{\mathrm{i}(k_1+k_2+k_3)\tau_0 } \right] \, .
%  \sum_{i=1}^{3} k_i
\eea

In the squeezed limit with $k_1 \ll k_2=k_3 \equiv k$ and $k<k_0$ this expression reduces to the following analytical formula
\bea\label{fnlslb}
F_{NL}^{<SL}(k_1,k) \approx -\frac{5}{6}\frac{H^5}{(2\pi)^4P_{\zeta}^2(k_*)} \frac{\phi_a^{+2}a(\tau_e)}{(\lambda^+)^3 (\phi_b^+)^6} \frac{1}{k} \Biggl[\frac{2k+k_1}{k_0}\cos{\left(\frac{2k+k_1}{k_0}\right)}\\ \nonumber
+\left(\frac{k}{k_0}\frac{2k_1+k}{k_0}-1\right)\sin{\left(\frac{2k+k_1}{k_0}\right)} \Biggr]\,.
\eea
As shown in fig.~(\ref{FNLfnlslb}) the analytical approximation is in good agreement with the numerical results. Here and in any other approximation for the $F_{NL}$ as defined in eq.~\eqn{FNL} we use eq.~\eqn{mps} for the spectrum $P_{\zeta}$. 

In the large scale equilateral limit, when  $k_1 = k_2=k_3 \equiv k \ll k_0 $ eq.~\eqn{fnlsl}  becomes
\bea\label{fnlelb}
F_{NL}^{<EL}(k) \approx -\frac{5}{6}\frac{H^5}{(2\pi)^4P_{\zeta}^2(k_*)} \frac{\phi_a^{+2}a(\tau_e)}{(\lambda^+)^3 (\phi_b^+)^6} \frac{1}{k} \left[\frac{3k}{k_0}\cos{\left(\frac{3k}{k_0}\right)}+\left(\frac{3k^2}{k_0^2}-1\right)\sin{\left(\frac{3k}{k_0}\right)} \right]\,.
\eea
Numerical result and eq.~\eqn{fnlelb} are in good agreement as shown in fig.~(\ref{FNLfnlelb}).
In figs.~(\ref{FNLfnlslb}-\ref{FNLfnlelb}) we have evaluated both the numerical and analytical expressions at a time $\tau_e$ corresponding approximately to $10e$-folds after the feature \cite{a1,a2,bingo,numerical2013}. As can be seen in figs.~(\ref{FNLfnlslb}-\ref{FNLfnlelb}) the large scale bispectrum in the squeezed and equilateral limits have a very similar form and are linearly suppressed.
%*********************************************************************************************************************************************
\subsection{Small scales}
In the small scale limit, when $k_i > k_0$, it is convenient to re-write the expression for $F_{NL}$ as
\bea \label{FNLss}
F_{NL}^>(k_1,k_2,k_3) \approx \frac{20}{3(2\pi)^4}\frac{(k_1 k_2 k_3)^3}{k_1^3+k_2^3+k_3^3}\frac{1}{P_{\zeta}^2(k_*)} \Im \biggl[ \zeta(\tau_e,k_1) \zeta(\tau_e,k_2)\zeta(\tau_e,k_3) \biggl( 2 I_1 (k_1,k_2,k_3)\\ \nonumber
 - k_1^2 I_2 (k_1,k_2,k_3) \biggr)  + \mbox{ two permutations of } k_1, k_2,\mbox{ and } k_3  \biggr] \, ,
\eea
where
\bea \label{integral1}
 I_1 (k_1,k_2,k_3) \equiv \int^{\tau_e}_{\tau_0} d\tau \, \eta(\tau) \epsilon(\tau) a(\tau)^2 \zeta^*(\tau,k_1) \zeta'{}^*(\tau,k_2)\zeta'{}^*(\tau,k_3) \\ \nonumber 
\approx \int^{\tau_e}_{\tau_0} d\tau \Bigl[ \lambda^+ (\lambda^-)^2  \phi_a^+ \phi_a^- a(\tau)^{2+ \lambda^-} + (\lambda^-)^3 (\phi_a^-)^2 a(\tau)^{2+2\lambda^-}  \Bigr] \zeta^*(\tau,k_1) \zeta'{}^*(\tau,k_2)\zeta'{}^*(\tau,k_3) \, ,\\
\label{integral2}
   I_2 (k_1,k_2,k_3) \equiv \int^{\tau_e}_{\tau_0} d\tau \, \eta(\tau) \epsilon(\tau) a(\tau)^2 \zeta^*(\tau,k_1) \zeta^*(\tau,k_2)\zeta^*(\tau,k_3) \\ \nonumber
  = \int^{\tau_e}_{\tau_0} d\tau \Bigl[ \lambda^+ (\lambda^-)^2  \phi_a^+ \phi_a^- a(\tau)^{2+ \lambda^-}
+ (\lambda^-)^3 (\phi_a^-)^2 a(\tau)^{2+2\lambda^-}  \Bigr] \zeta^*(\tau,k_1) \zeta^*(\tau,k_2)\zeta^*(\tau,k_3) \, .
\eea
In the above expressions we have used  eq.~\eqn{slowrollapprox} to derive an approximation for $\eta \epsilon a^2$ as
\bea \label{slowrollapprox2}
  \eta \epsilon a^2 = \left[(\lambda^+)^2 \phi_a^+ a^{\lambda^+} + (\lambda^-)^2\phi_a^- a^{\lambda^-}  \right] \left[\lambda^+ \phi_a^+ a^{\lambda^+}+ \lambda^- \phi_a^- a^{\lambda^-} \right] a^2 \\ \nonumber 
  = \left[ (\lambda^+)^3 (\phi_a^+)^2 a^{2\lambda^+} + (\lambda^+ +\lambda^-) \lambda^+ \lambda^-  \phi_a^+ \phi_a^- a^{\lambda^+ + \lambda^-} + (\lambda^-)^3 (\phi_a^-)^2 a^{2\lambda^-} \right] a^2 \\ \nonumber 
  \approx \lambda^+ (\lambda^-)^2  \phi_a^+ \phi_a^- a^{2+ \lambda^-} + (\lambda^-)^3 (\phi_a^-)^2 a^{2+2\lambda^-} \, .
\eea

All the integrals we need to compute have a similar form, so it is useful to re-write eqs. \eqn{integral1} and \eqn{integral2} as	
\bea \label{integral12}
I_i (k_1,k_2,k_3) \equiv \biggl[\lambda^+ (\lambda^-)^2  \phi_a^+ \phi_a^- \mathcal{A}_i(\tau,k_1,k_2,k_3,q_1)
 + (\lambda^-)^3 (\phi_a^-)^2 \mathcal{A}_i(\tau,k_1,k_2,k_3,q_2) \biggr] \Bigg|^{\tau_e}_{\tau_0} \\ \nonumber
\approx \lambda^+ (\lambda^-)^2  \phi_a^+ \phi_a^- \mathcal{A}_i(\tau_0,k_1,k_2,k_3,q_1) 
+ (\lambda^-)^3 (\phi_a^-)^2 \mathcal{A}_i(\tau_0,k_1,k_2,k_3,q_2) 
\, ,
\eea
where we have defined
\bea
\mathcal{A}_1(\tau,k_1,k_2,k_3,q) &\equiv& \int d\tau \, a(\tau)^q \zeta^*(\tau,k_1) \zeta'{}^*(\tau,k_2)\zeta'{}^*(\tau,k_3), \\
\mathcal{A}_2(\tau,k_1,k_2,k_3,q) &\equiv& \int d\tau \, a(\tau)^q \zeta^*(\tau,k_1) \zeta^*(\tau,k_2)\zeta^*(\tau,k_3)\,, \\
q_1=2+\lambda^- &,&  q_2 =2+2\lambda^- \,.
\eea

The above integrals can be computed analytically in terms of $\Gamma$ functions and are given in details in Appendix A.

It is now possible to obtain a fully analytical template in the squeeze limit, when $k_2=k_3$, and $k_2 \gg k_1>k_0$ 
\bea \label{fnlsla}
  F_{NL}^>(k_1,k_2)\approx \frac{20}{3(2\pi)^4}\frac{(k_1 k_2)^3}{P_{\zeta}^2(k_*)} \Im \biggl[ \zeta(\tau_e,k_1) \zeta(\tau_e,k_2)^2 \biggl( I_1(k_1,k_2,k_2) \\ \nonumber
  + 2 I_1(k_2,k_1,k_2)-k^2 I_2(k_1,k_2,k_2)  \biggr) \biggr] \, .
\eea
In the equilateral limit, when $k\equiv k_1=k_2=k_3$ and $k>k_0$, instead  we have
\bea \label{fnlela}
  F_{NL}^>(k_1)\approx \frac{20}{3(2\pi)^4}\frac{k^6}{P_{\zeta}^2(k_*)}\Im \biggl[ \zeta(\tau_e,k)^3 \Bigl( 3 I_1(k) -k^2 I_2(k) \Bigr) \biggr] \, .
\eea
% In the above template $\delta$ is a phase shift parameter which varies for different models or limits.
% and has to be determined case by case through a fitting procedure with the numerical or observational results. This is due to the fact that the analytical approximation we use for the modes is  good for the absolute value of curvature perturbations $|\zeta|$, but it does not always approximate well the phase. 
%For the same reason a similar fitting procedure was used for the bispectrum  in \cite{aer}.
Numerical results and the analytical templates are in good agreement both in the squeezed and equilateral limits as shown in figs.~(\ref{FNLfnlsla}-\ref{FNLfnlela}). In the squeezed and equilateral small scale limits the bispectrum has an oscillatory behavior whose phase and amplitude depend on the value of the parameters $n$ and $\lambda$  as it can be seen in figs.~(\ref{FNLfnlsla}-\ref{FNLfnlela}).
The amplitude is inversely proportional to the scale as show in figs.~(\ref{FNLfnlsla}-\ref{FNLfnlela}).
As previously observed, all the results derived can be trusted only up to cut-off scales beyond which the Heaviside approximation  is not valid as discussed in more details in \cite{aer,pp}. The same applies to other similar models previously studied such as the well known Starobinsky model \cite{starobinsky}.

\begin{figure}
 \begin{minipage}{.45\textwidth}
  \includegraphics[scale=0.6]{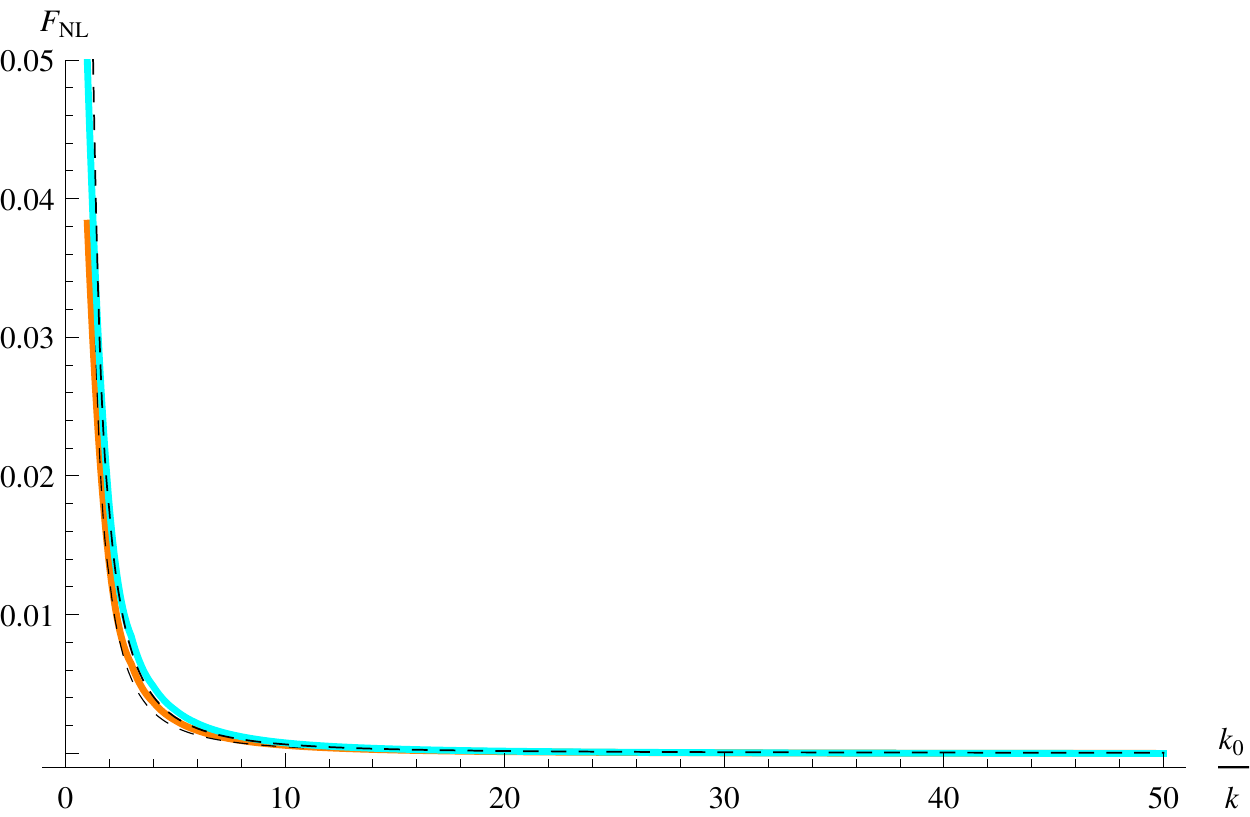}
 \end{minipage}
  \begin{minipage}{.45\textwidth}
 \includegraphics[scale=0.6]{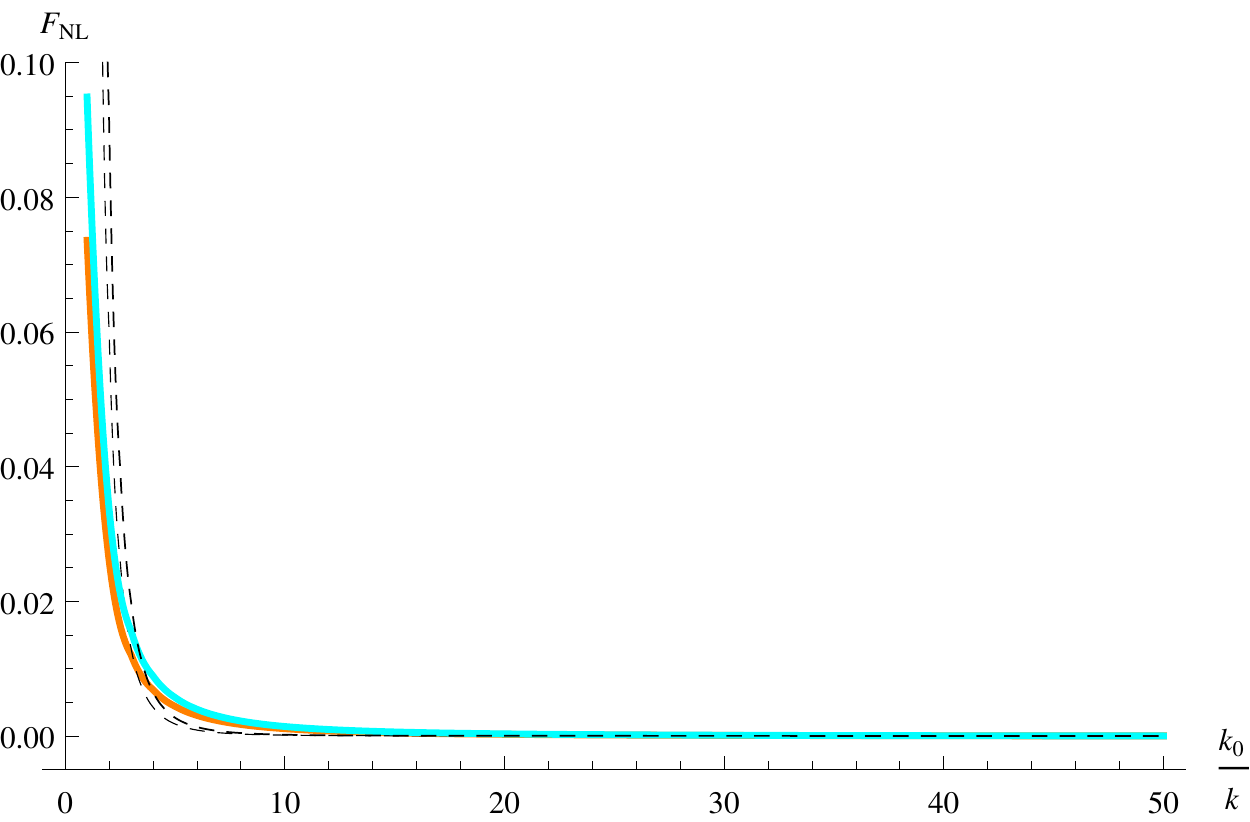}
  \end{minipage}
 \begin{minipage}{.45\textwidth}
  \includegraphics[scale=0.6]{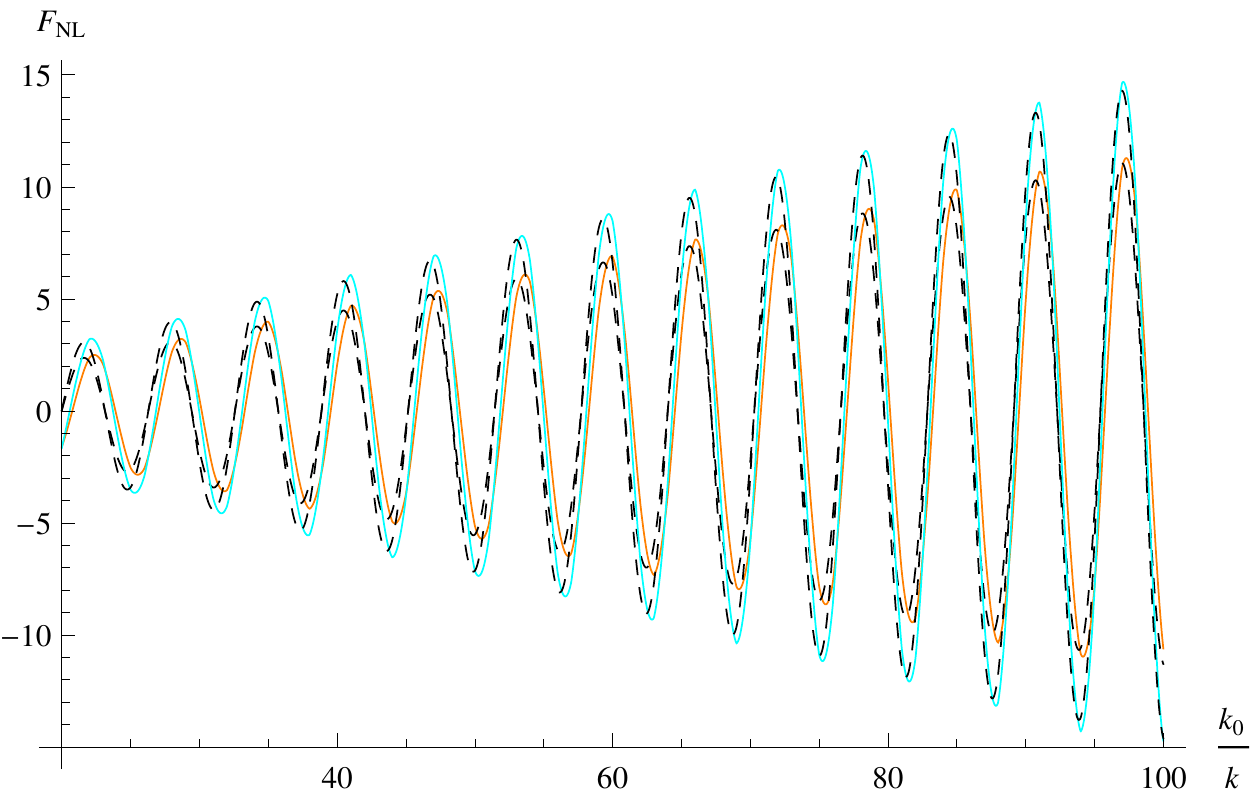}
 \end{minipage}
 \begin{minipage}{.45\textwidth}
  \includegraphics[scale=0.6]{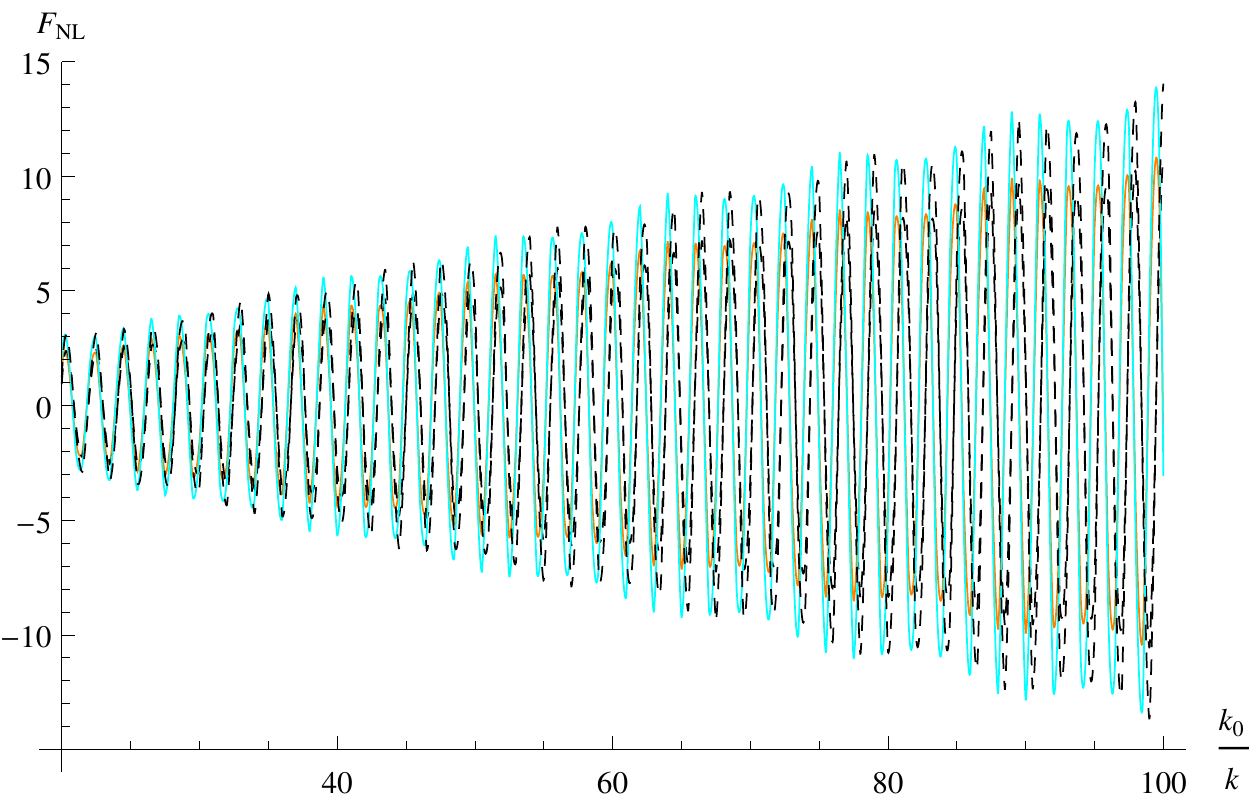}
 \end{minipage}
 \caption{From left to right and top to bottom and with $n=3$, $\lambda=-8\times10^{-20}$ (orange) and $\lambda=-6\times10^{-20}$ (cyan). The squeezed limit for large scales $F_{NL}(k_0/500,k,k)$. The equilateral limit for large scales $F_{NL}(k,k,k)$. The squeezed limit for small scales $F_{NL}(k,1000k_0,1000k_0)$. The equilateral limit for small scales $F_{NL}(k,k,k)$. The dashed lines are the analytical approximations. This choice of parameters is the same used in fig.(9), for models able to account for the  observed large scale suppression.}
\label{fit}
\end{figure}

\subsection{Behavior of the small scales bispectrum}
As seen in figs.~(\ref{FNLfnlsla}-\ref{FNLfnlela}) both the equilateral and the squeezed limit small scale bispectrum do not behave in the same way as the spectrum and the slow parameters respect to the variation of $n$.
To clarify this we can write the bispectrum $B_{\zeta}$ in eq.~\eqn{b} as
\bea 
\label{prod}
B_{\zeta} \propto \Im(B_1 B_2)=\Re(B_1)\Im(B_2) + \Im(B_1)\Re(B_2) \,, 
\eea
where 
\bea
B_1&=&\zeta(\tau_e,k_1)\zeta(\tau_e,k_2)\zeta(\tau_e,k_3) \,, \\
B_2&=& 2 I_1 (k_1,k_2,k_3)- k_1^2 I_2 (k_1,k_2,k_3) + \mbox{ two permutations of } k_1, k_2,\mbox{ and } k_3 \, .
\eea
First of all we can see from fig.~(17) and fig.~(20) that the dominant contribution to the bispectrum comes from the term $\Im(B_1)\Re(B_2)$, which is in fact behaving in the same way as the bispectrum respect to the variaton of $n$. 
Both the terms $\Re(B_2)$ and $\Im(B_1)$ are behaving like the spectrum, i.e., are larger for larger values of $n$, but since  $\Im(B_1)$ is negative, their product $\Im(B_1)\Re(B_2)$, and consequently the bispectrum which is dominated by it, is behaving in the opposite way, i.e. is decreasing when $n$ in increasing. The effect is not noticeable in the case of $n=2/3$, because in this case  $\Re(B_2)$ is very closed to zero, while it is clear for $n=3$ and $n=4$.

%While $\Re{B_2}$ is 
%\textcolor{red} {We use the analytic results to show %the behavior of $B_1$ and $B_2$ in figs.~(\ref{B1}-%\ref{reim}).}

\begin{figure}
 \begin{minipage}{.45\textwidth}
  \includegraphics[scale=0.6]{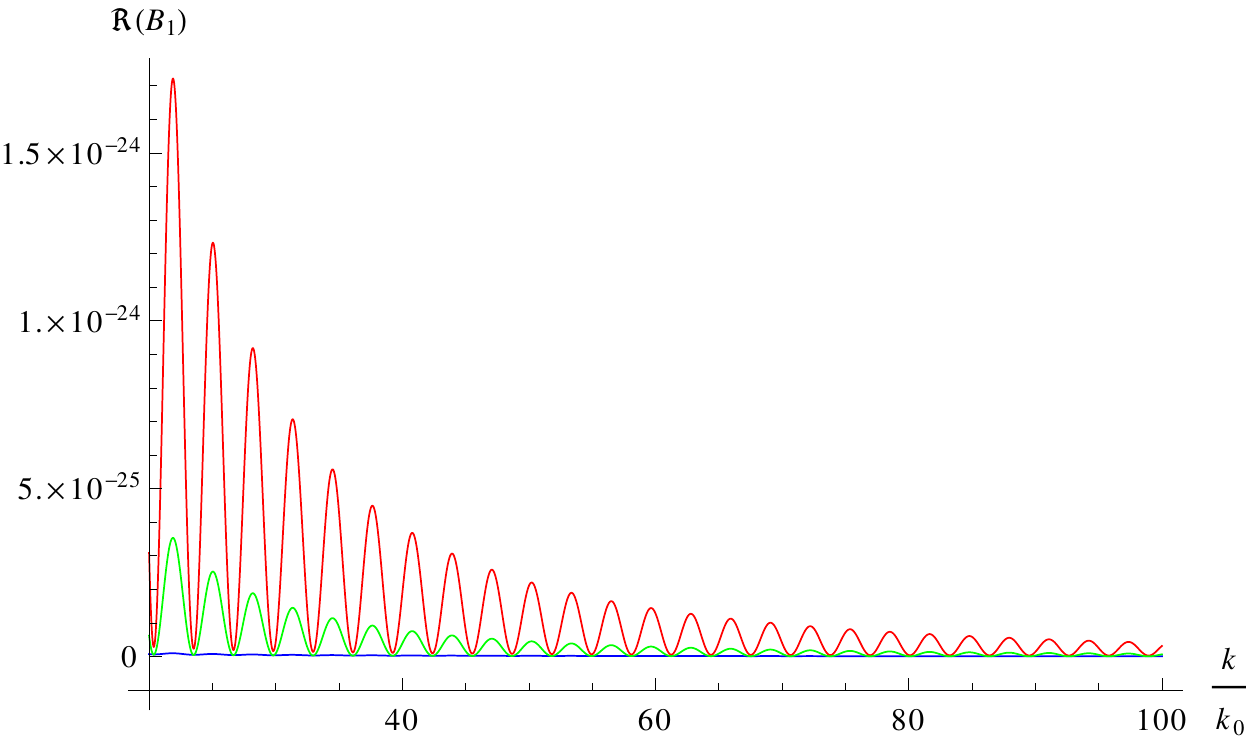}
  \end{minipage}
 \begin{minipage}{.45\textwidth}
  \includegraphics[scale=0.6]{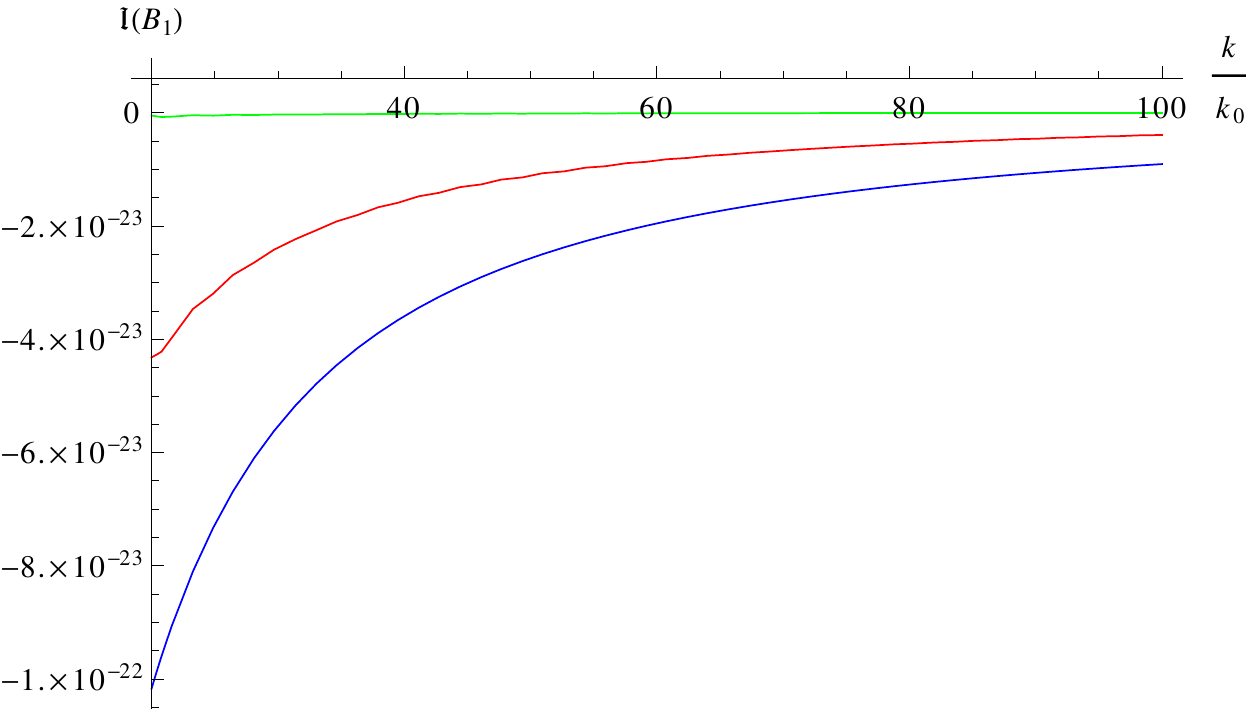}
 \end{minipage}
 \caption{The real part of $B_1$ is plotted on the left, and the imaginary part on the right, for the small scales  squeezed limit. The parameter   $\lambda$ is constant, $\lambda=3.9\times10^{-19}$, while $n=2/3$ (blue), $n=3$ (red), and $n=4$ (green).}
\label{B1}
\end{figure}

\begin{figure}
 \begin{minipage}{.45\textwidth}
  \includegraphics[scale=0.6]{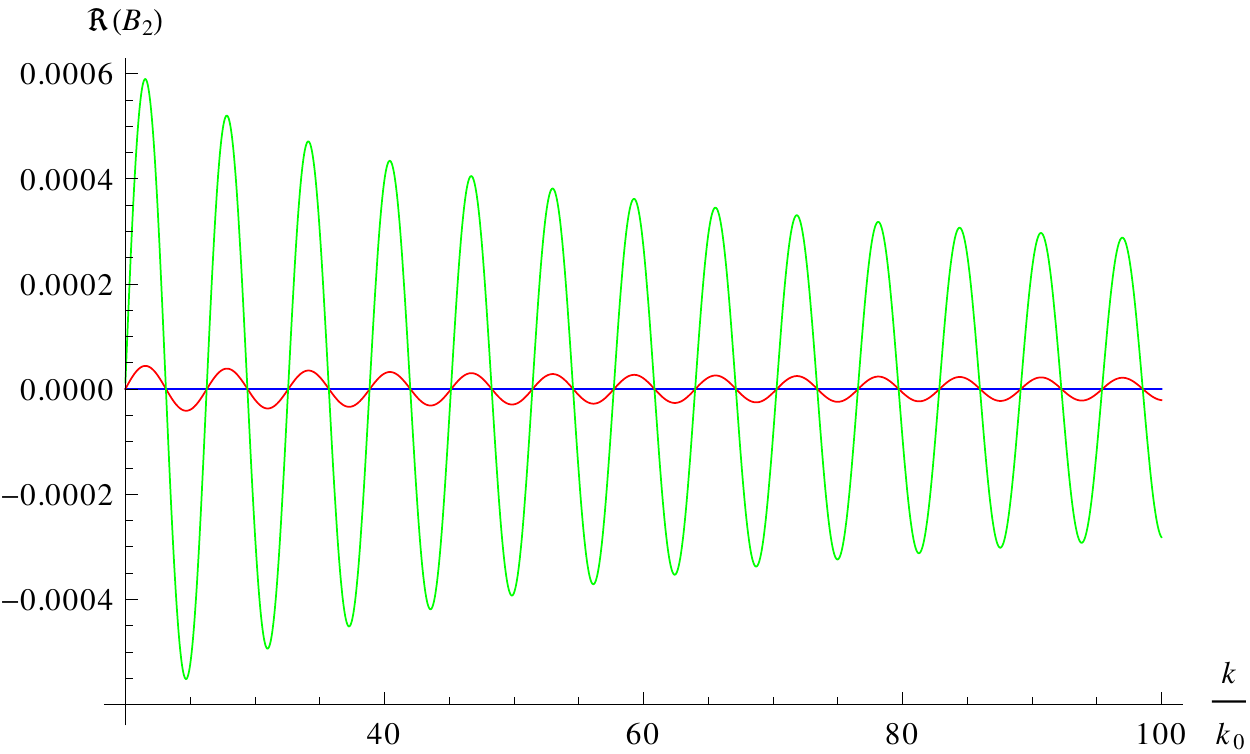}
  \end{minipage}
 \begin{minipage}{.45\textwidth}
  \includegraphics[scale=0.6]{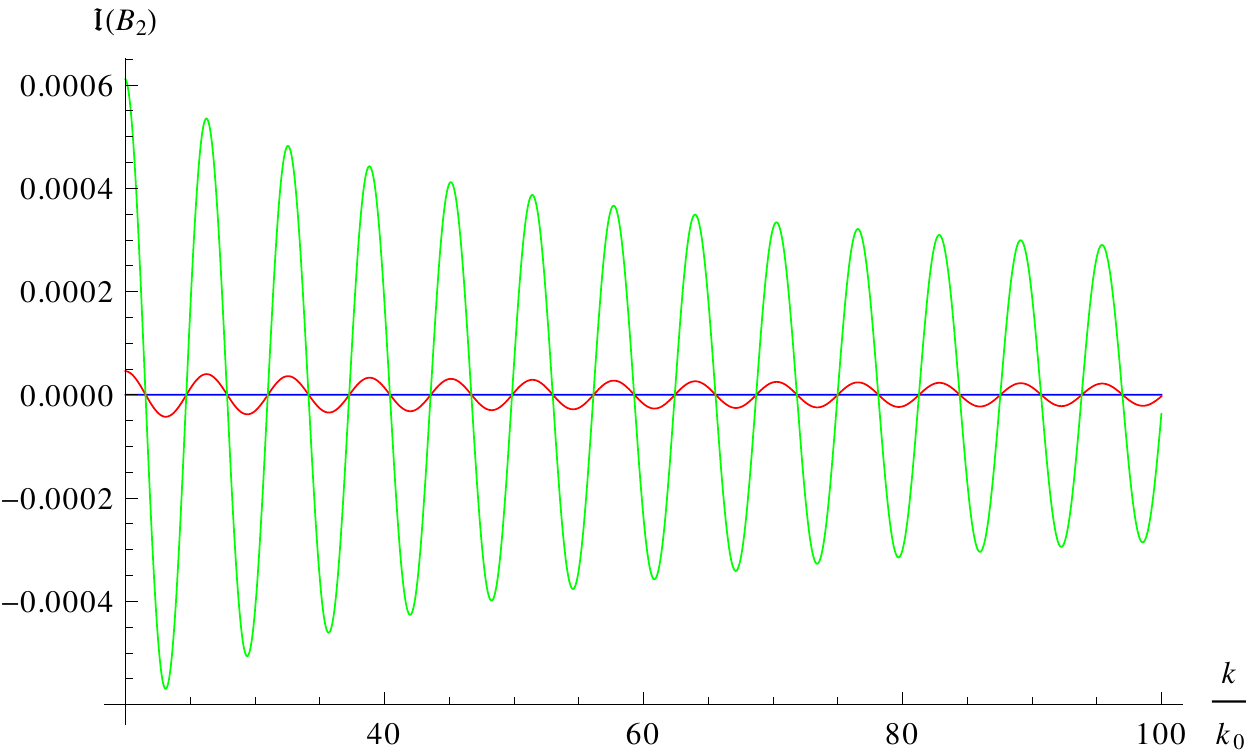}
 \end{minipage}
 \caption{The real part of $B_2$ is plotted on the left, and the imaginary part on the right, for the small scales  squeezed limit. The parameter   $\lambda$ is constant, $\lambda=3.9\times10^{-19}$, while $n=2/3$ (blue), $n=3$ (red), and $n=4$ (green).}
\label{B2}
\end{figure}

\begin{figure}
 \begin{minipage}{.45\textwidth}
  \includegraphics[scale=0.6]{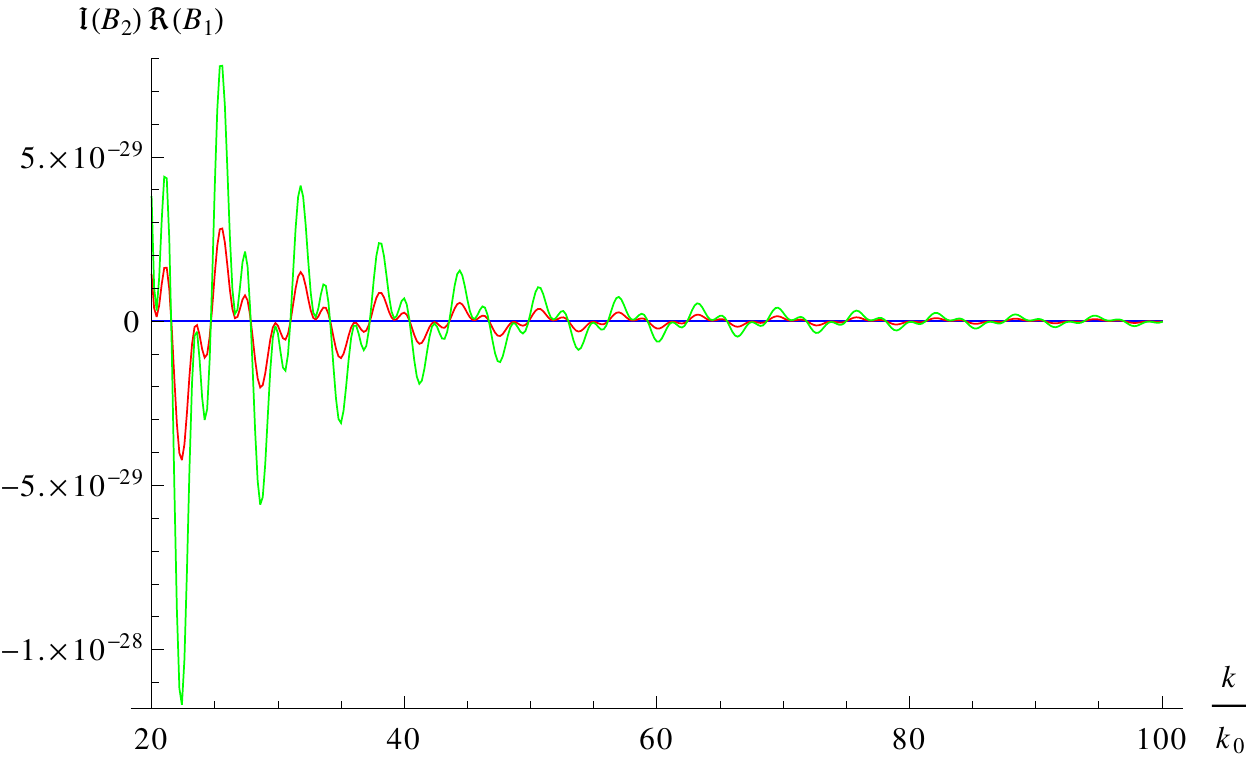}
  \end{minipage}
 \begin{minipage}{.45\textwidth}
  \includegraphics[scale=0.6]{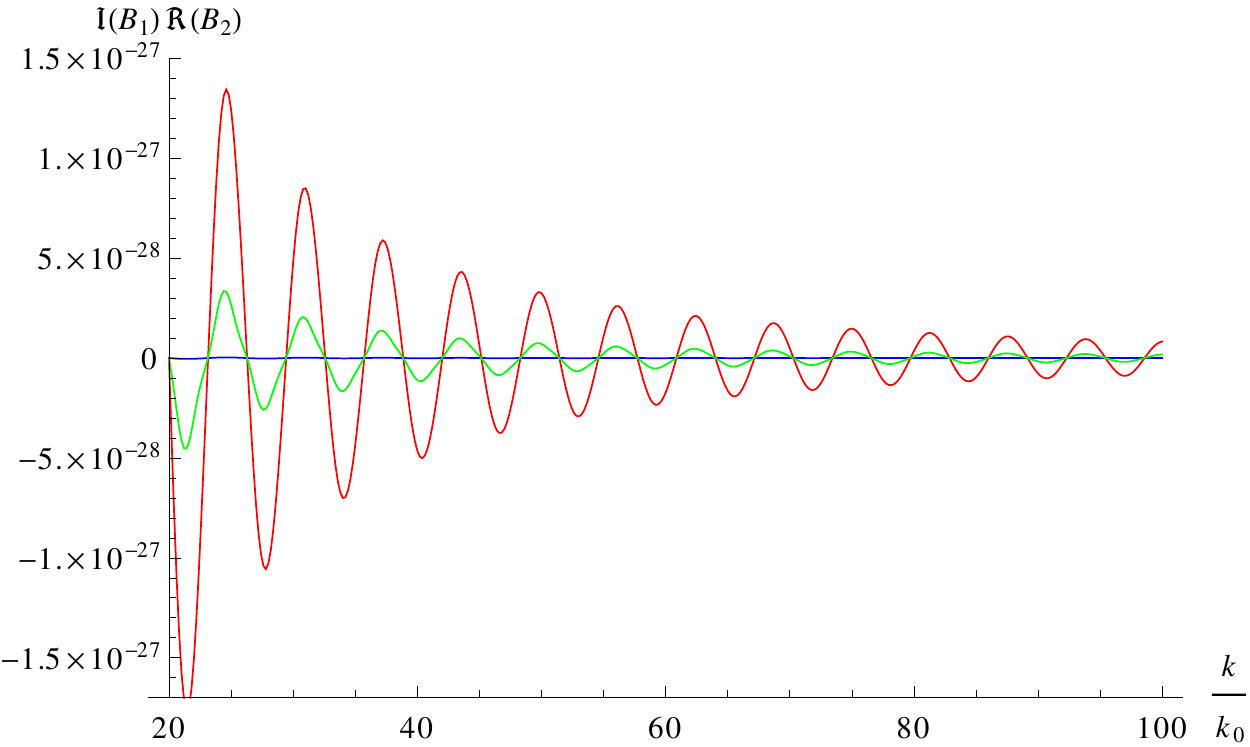}
 \end{minipage}
 \caption{
 The products $\Re(B_1)\Im(B_2)$ and $ \Im(B_1)\Re(B_2)$ are plotted for the small scales  squeezed limit. The parameter   $\lambda$ is constant, $\lambda=3.9\times10^{-19}$, while $n=2/3$ (blue), $n=3$ (red), and $n=4$ (green).}
\label{reim}
\end{figure}
%************************
\begin{figure}
 \begin{minipage}{.45\textwidth}
  \includegraphics[scale=0.6]{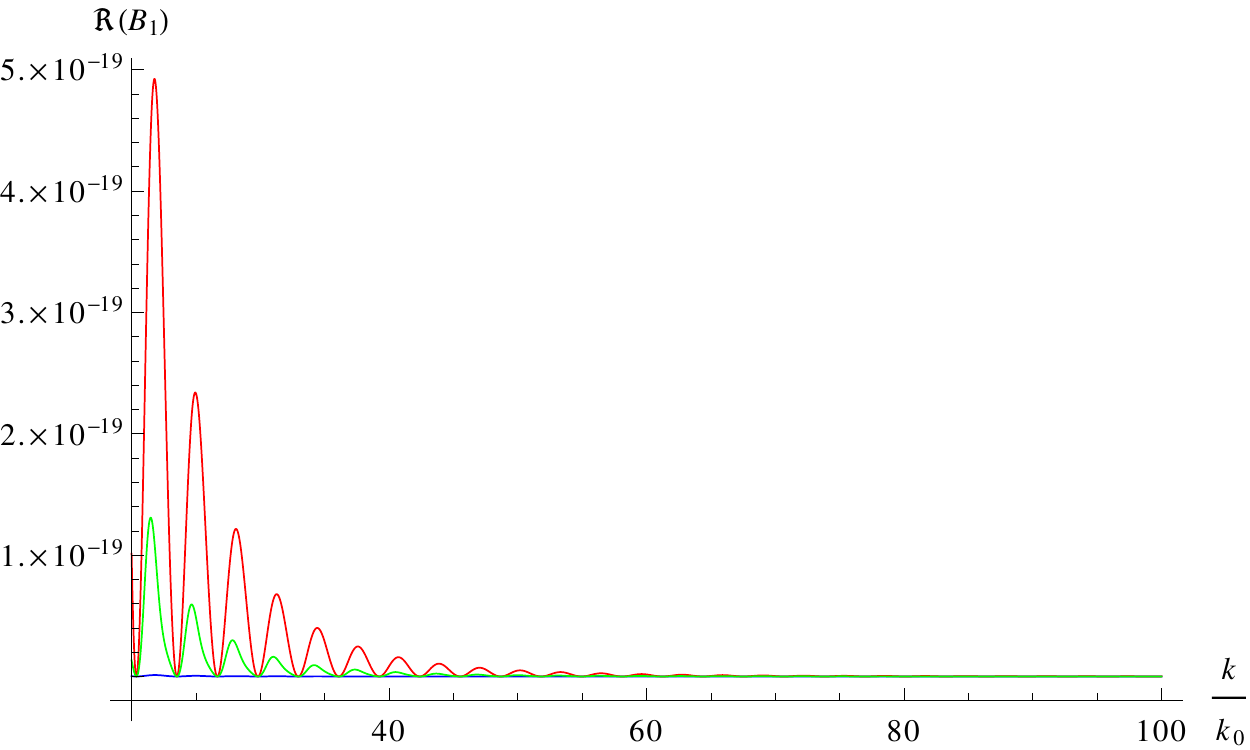}
  \end{minipage}
 \begin{minipage}{.45\textwidth}
  \includegraphics[scale=0.6]{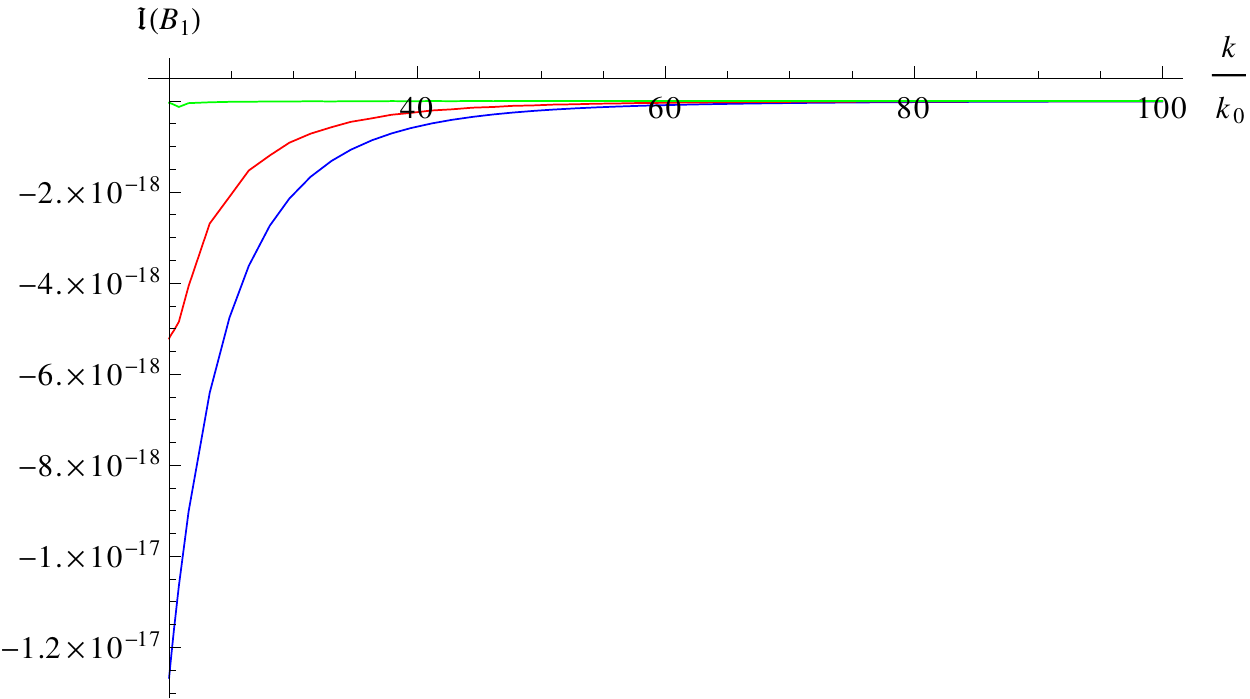}
 \end{minipage}
 \caption{The real part of $B_1$ is plotted on the left, and the imaginary part on the right, for the small scales  equilateral limit. The parameter   $\lambda$ is constant, $\lambda=3.9\times10^{-19}$, while $n=2/3$ (blue), $n=3$ (red), and $n=4$ (green).}
\label{B1el}
\end{figure}

\begin{figure}
 \begin{minipage}{.45\textwidth}
  \includegraphics[scale=0.6]{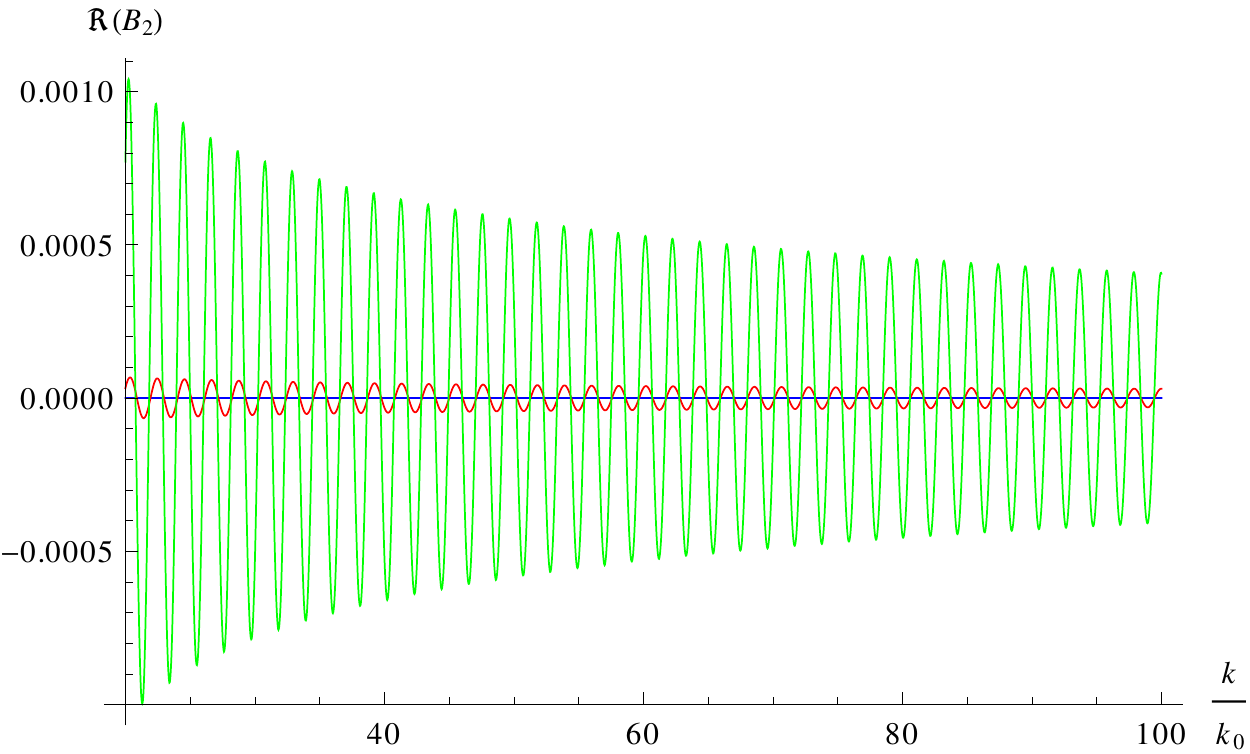}
  \end{minipage}
 \begin{minipage}{.45\textwidth}
  \includegraphics[scale=0.6]{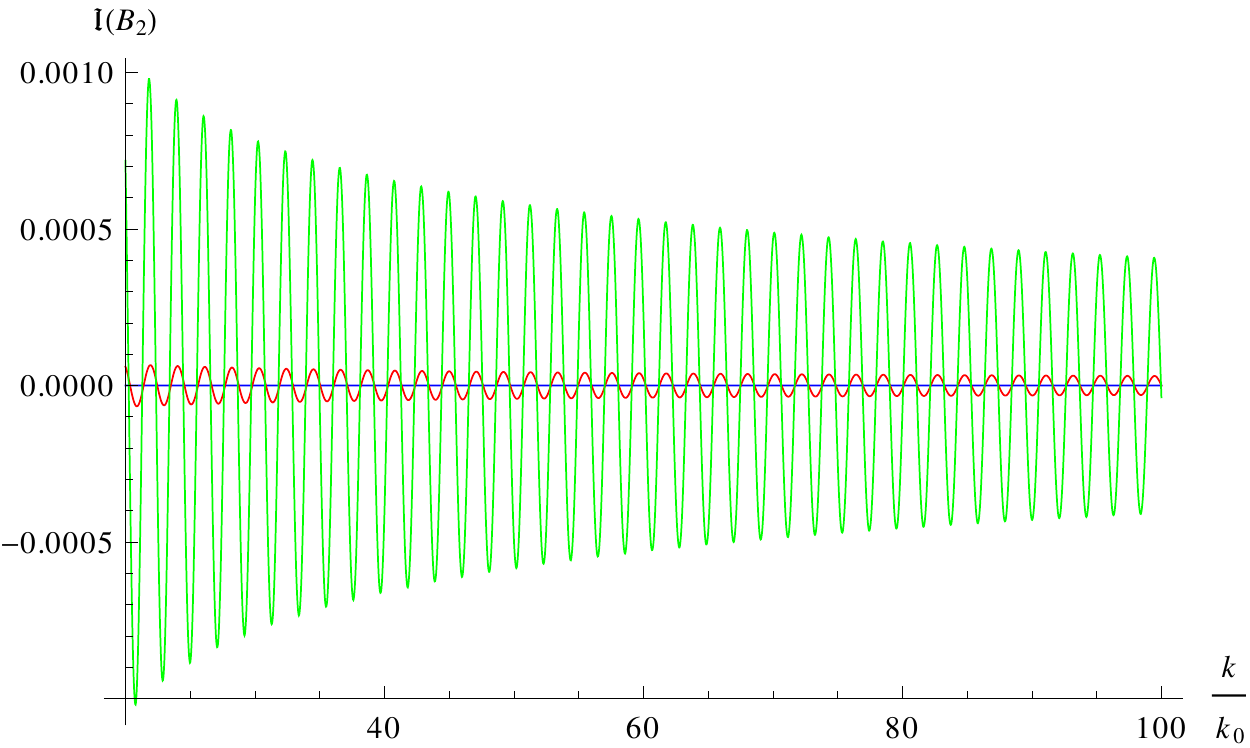}
 \end{minipage}
 \caption{The real part of $B_2$ is plotted on the left, and the imaginary part on the right, for the small scales equilateral limit. The parameter   $\lambda$ is constant, $\lambda=3.9\times10^{-19}$, while $n=2/3$ (blue), $n=3$ (red), and $n=4$ (green).}
\label{B2el}
\end{figure}

\begin{figure}
 \begin{minipage}{.45\textwidth}
  \includegraphics[scale=0.6]{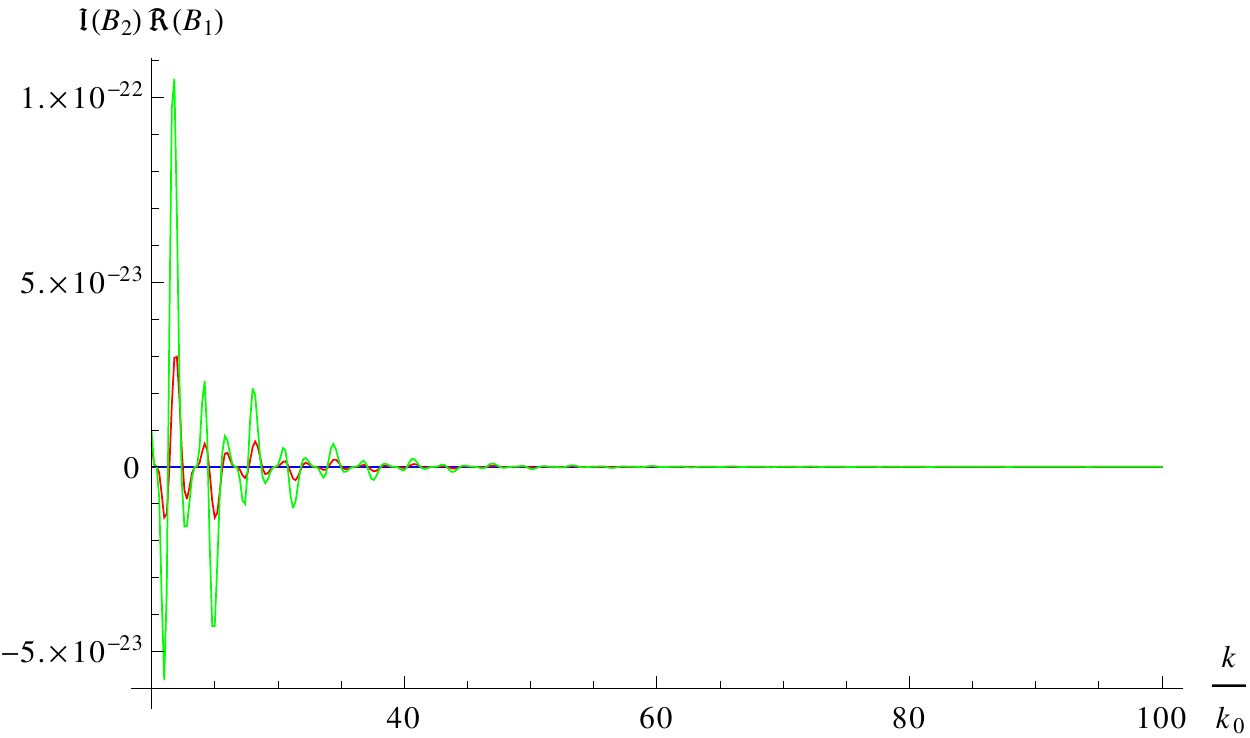}
  \end{minipage}
 \begin{minipage}{.45\textwidth}
  \includegraphics[scale=0.6]{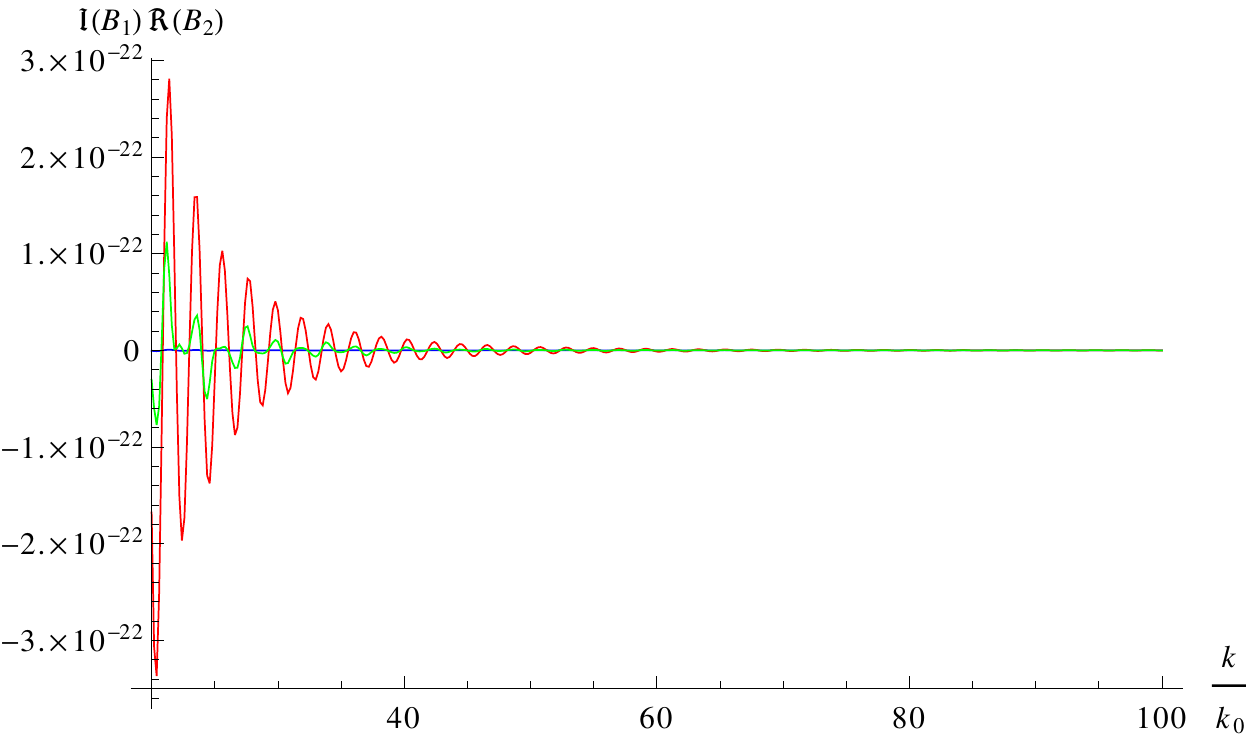}
 \end{minipage}
 \caption{
 The products $\Re(B_1)\Im(B_2)$ and $ \Im(B_1)\Re(B_2)$ are plotted for the small scales  squeezed limit. The parameter   $\lambda$ is constant, $\lambda=3.9\times10^{-19}$, while $n=2/3$ (blue), $n=3$ (red), and $n=4$ (green).}
\label{reimel}
\end{figure}
%***********************************************************************************************************************************
\section{Conclusions}
We have studied the effects of a general type of features produced by  discontinuities of the derivatives of the potential. We found that each different type of feature  has distinctive effects on the spectrum and bispectrum of curvature perturbations which depend both on the order $n$ and on the amplitude $\lambda$ of discontinuity. The spectrum of primordial curvature perturbations shows oscillations around the scale $k_0$ which leaves the horizon at the time $\tau_0$ when the feature occurs, with amplitude and phase determined by the parameters $n$ and $\lambda$.

Both in the squeezed and equilateral small scale limit the bispectrum has an oscillatory behavior whose phase depends on the parameters determining the discontinuity, and whose amplitude is inversely proportional to the scale. The large scale bispectrum in the squeezed and equilateral limits have a very similar form and are linearly suppressed.

The analytical approximation for the spectrum is in good agreement with the numerical results, and improves substantially the accuracy for large scales respect to previous studies. The analytical approximations for the bispectrum are in good agreement with numerical calculations  at large scales in both the squeeze and equilateral limit.
At small scales we found an analytical template which is in very good agreement with the numerical calculations both in the squeezed and equilateral limit, and is able to account for both the oscillations and the amplitude of the bispectrum.

The type of feature we have studied generalize previous models such as the Starobinsky model or the mass step \cite{aer}, providing a general framework to classify and model phenomenologically non Gaussian features in CMB observations or in large scale structure survey data. In the future it would be interesting to find the parameters which better fit different non Gaussian features in observational data and to investigate what more fundamental physical mechanism, such as phase transitions for example, could actually produce these features.
%It would also be interesting to study the effects of these generalized features on tensorial perturbations, with special attention to the large  tensor to scalar ratio which was recently claimed to have been observed  by the experiment $BICEP2$ \cite{bicep2}.

\acknowledgments
%The authors thank the anonymous referee for useful comments to improve the the manuscript.
AER work was  supported by the Dedicacion exclusiva and Sostenibilidad programs
at UDEA, the UDEA CODI projects IN10219CE and 2015-4044.
This work was supported by the European Union (European Social Fund, ESF) and Greek national funds under the “ARISTEIA II” Action. 

%, the Dedicacion exclusica and Sostenibilidad programs at UDEA, the UDEA CODI
%project IN10219CE.

\begin{appendix}
 \section{} \label{A1}
In this appendix we obtain analytical approximations for the integrals which are necessary for the calculation of small scale limit bispectrum
\bea
\mathcal{A}_1(\tau,k_1,k_2,k_3,q) &\equiv& \int d\tau \, a(\tau)^q \zeta^*(\tau,k_1) \zeta'{}^*(\tau,k_2)\zeta'{}^*(\tau,k_3) \, , \\
\mathcal{A}_2(\tau,k_1,k_2,k_3,q) &\equiv& \int d\tau \, a(\tau)^q \zeta^*(\tau,k_1) \zeta^*(\tau,k_2)\zeta^*(\tau,k_3)\,.
\eea
In order to simplify the calculation we fix $\epsilon(\tau)=\epsilon(\tau_0)$ only in the analytical approximation for perturbations modes in eq.~(\ref{alphabeta}), while we keep $\epsilon(\tau),\eta(\tau)$ as  functions of conformal time when they appear explicitly in the integrand.

After some rather cumbersome calculation the final result can be written in this general form
\bea\label{antiderivatives}
&&\mathcal{A}_i(\tau,k_1,k_2,k_3,q)= \frac{(-1)^i(k_2 k_3)^{2(2-i)} H_0^{3-q}}{\left(4 \epsilon_0 k_1 k_2 k_3\right)^{3/2}} \times \\ \nonumber
\Biggl\{ 
\alpha^*_{k_1} \biggl[ && \alpha^*_{k_2} \Bigl( \mathcal{B}_i(\tau,k_1,k_2,k_3,q) \alpha^*_{k_3} - \mathcal{B}_i(\tau,k_1,k_2,-k_3,q) \beta^*_{k_3} \Bigr) \\ \nonumber
+ && \beta^*_{k_2} \Bigl( -\mathcal{B}_i(\tau,k_1,-k_2,k_3,q) \alpha^*_{k_3} + \mathcal{B}_i(\tau,k_1,-k_2,-k_3,q) \beta^*_{k_3} \Bigr) \biggr] \\ \nonumber
+ \beta^*_{k_1} \biggl[ && \beta^*_{k_2} \Bigl( \mathcal{B}^*_i(\tau,k_1,k_2,k_3,q) \beta^*_{k_3} - \mathcal{B}^*_i(\tau,k_1,k_2,-k_3,q) \alpha^*_{k_3} \Bigr) \\ \nonumber
+ && \alpha^*_{k_2} \Bigl( -\mathcal{B}^*_i(\tau,k_1,-k_2,k_3,q) \beta^*_{k_3} + \mathcal{B}^*_i(\tau,k_1,-k_2,-k_3,q) \alpha^*_{k_3} \Bigr) \biggr]
\Biggr\} ,
\eea
where
\bea
&& \mathcal{B}_1=(\mathrm{i}k_T )^{q-4} \Bigl(k_T \Gamma (3-q,-\mathrm{i} \tau k_T ) +k_1 \Gamma (4-q,-\mathrm{i}\tau k_T ) \Bigr)  \, , \\ \nonumber
&& \mathcal{B}_2= (\mathrm{i}k_T )^{q-4} \Bigl[ k_T^3 \Bigl( \Gamma (1-q,-\mathrm{i} \tau k_T ) + \Gamma (2-q,-\mathrm{i} \tau k_T ) \Bigr) \\ 
&&+ k_T \sum_{i \ne j}^{3}k_ik_j \Gamma (3-q,-\mathrm{i} \tau k_T ) +k_1 \Gamma (4-q,-\mathrm{i} \tau k_T ) \Bigr] \, , \\
&&k_T=k_1+k_2+k_3 \,, \nonumber
\eea
 and the $\Gamma$ denotes the incomplete gamma functions defined by
\bea
\Gamma(r,x)= \int_x^{\infty} t^{r-1} e^{-t} dt  \,.
\eea

\end{appendix}

\bibliography{Bibliography}
\bibliographystyle{h-physrev4}
\end{document}